\documentclass[aps,prb,superscriptaddress,preprint]{revtex4}

\usepackage{physics}
\usepackage{siunitx}

\usepackage{amssymb}
\usepackage{amsmath}
\usepackage{amsfonts}
\usepackage{graphicx}
\usepackage{subfigure}

\usepackage{color}
\usepackage[utf8]{inputenc}
\usepackage{multirow}
\usepackage{url}
\usepackage{placeins}
\usepackage{diagbox}

\begin{document}

\newcounter{SOMsecCounter}
\setcounter{SOMsecCounter}{0}
\refstepcounter{SOMsecCounter} 
\label{Sec:SOM_Bfield}
\refstepcounter{SOMsecCounter} 
\label{Sec:SOM_Dip}
\refstepcounter{SOMsecCounter} 
\label{Sec:SOM_SOC}
\refstepcounter{SOMsecCounter} 
\label{Sec:SOM_Spin}
\refstepcounter{SOMsecCounter} 
\label{Sec:SOM_TDM}
\refstepcounter{SOMsecCounter} 
\label{Sec:SOM_Jones}

\makeatletter
\renewcommand\@biblabel[1]{#1.}
\makeatother

\title{Evolution of atomic optical selection rules upon gradual symmetry lowering}

\author{G.~J.~J.~Lof}
\affiliation{Zernike Institute for Advanced Materials,
University of Groningen, The Netherlands}
\author{C.~H.~van~der~Wal}
\affiliation{Zernike Institute for Advanced Materials,
University of Groningen, The Netherlands}
\author{R.~W.~A.~Havenith}
\affiliation{Zernike Institute for Advanced Materials,
	University of Groningen, The Netherlands}
\affiliation{Stratingh Institute for Chemistry, University of Groningen, The Netherlands}
\affiliation{Ghent Quantum Chemistry Group, Department of Inorganic and Physical Chemistry, Ghent University, Belgium}


\date{\today}


\begin{abstract}
	For atoms and crystals with an ideal symmetry, the optical selection
	rules for electronic transitions are well covered in physics textbooks.
	However, in studies of material systems one often encounters systems
	with a weakly distorted symmetry. Insight and intuition for how optical
	selection rules change when an ideal symmetry is gradually distorted is,
	nevertheless, little addressed in literature. We present here a
	detailed analysis of how a gradual symmetry distortion leads to a
	complete alteration of optical selection rules. As a model system, we
	consider the transitions between $1s$ and $2p$ sublevels of the hydrogen
	atom, which get distorted by placing charged particles in its
	environment. Upon increasing the distortion, part of the optical selection rules evolve from circular via elliptical to linear character, with an
	associated evolution between allowed and forbidden transitions. Our
	presentation combines an analytical approach with quantitative results
	from numerical simulations, thus providing insight in how the evolution
	occurs as a function of the strength of the distortion.
\end{abstract}

\maketitle

\section{Introduction}

A physical system is never completely isolated. Even in atomic
clocks\cite{grynberg2010introduction}, which use quantum oscillations in
atoms that are relatively insensitive to surrounding matter and fields,
the symmetry and dynamics of the quantum system of interest are affected
by the environment. Consequently, a system with ideal symmetry does not
exist, but the distortions due to the environment can be so weak that
their influence is not significant.

The symmetry of a system dictates its optical selection rules for
electronic transitions. Well-known behavior of such optical selection
rules is that absorbing circularly polarized light can orient the spin
(or, more generally, electronic angular momentum) of an electron that
gets
excited\cite{cohen1991quantum,foot2005atomsbook,atkins2011molecular}.
This occurs in systems of high symmetry, and is widely applied. A key
example is the use of alkali atoms (with spherical symmetry) such as
hydrogen, rubidium and cesium, for quantum optical studies and
technologies. A second important example is the optoelectronic control
in semiconductors with the tetrahedral zincblende lattice structure
(with GaAs as key example), where spintronic applications use spin
orientation by circularly polarized light\cite{fox2010optical}.

For other material systems, with a lower symmetry, optical
transitions couple more frequently purely to linearly polarized light.
This holds for excitonic transitions in most organic
molecules\cite{atkins2011molecular,dediu2009spinorganic}, and
transitions of molecule-like color centers in crystals, such as the
strongest transitions of the nitrogen-vacancy defect in
diamond\cite{chu2015nvcenter} (a widely-studied system for
quantum technologies).

There exist also many material systems which have a high but still
weakly distorted symmetry. For these cases it is much harder to assess
the optical selection rules with analytical methods, and this topic is
little covered in textbooks. Here detailed numerical calculations can
provide predictions, but it is much harder to obtain intuitive insight from
the output of such calculations.
Still, the elegance of computational physics and chemistry calculations
is that they can relatively easily reveal how the properties of matter
vary in dependence of parameter values.

In this work we provide a detailed theoretical analysis of how a gradual
symmetry distortion leads to a complete alteration of optical selection
rules. As a model system we use the hydrogen atom, and our results give
insight in how its optical selection rules (for transitions between a
$1s$ and $2p$ sublevel) change gradually for a gradual symmetry change
due to a disturbing environment.
The optical selection rules of the bare hydrogen atom can be described
analytically, and are well-known\cite{cohen1991quantum}. We use this as
a starting point.
We include in the discussion how they behave in a weak magnetic field,
since this is of interest for highlighting the  properties of the
selection rules.
We model the symmetry lowering due to an environment by placing the
hydrogen atom in a $C_{2v}$-symmetry arrangement of four negative point
charges, where the magnitude of the charges is varied. For this situation the
analytical calculations are too complicated, and we link the analysis to
numerical calculations of this system.
Our work thus also provides an interesting example of how modern methods
for numerically simulating matter can give insight in its properties at
a quantum mechanical level.

This manuscript is organized as follows. In Section~\ref{Sec:noSpin}, we
will introduce the bare hydrogen atom, first
without considering spin, and we focus on the electronic transitions
between the $1s$ and $2p$ states.
This serves as a summary of how this is treated in many textbooks on
atomic physics\cite{cohen1991quantum}, and for introducing the notations
we use. How a magnetic field affects these transitions is treated in
Supplementary Information Section~\ref{Sec:SOM_Bfield}
and~\ref{Sec:SOM_Dip}. Next, we expand this model in the usual manner by
also considering the electron spin and the effect of spin-orbit coupling
(SOC). This is presented in Section~\ref{Sec:withSpin} and
Supplementary Information Section~\ref{Sec:SOM_Spin}. In
Section~\ref{Sec:C2v}, we add the symmetry disturbance to the modeling,
by considering the hydrogen atom in a $C_{2v}$ arrangement of negative
point charges. For the analysis of this case we used numerical
simulation methods, that are also introduced in this section (we used
the CASSCF/RASSI\textendash SO method
\cite{roos2004effects,roos2004relativistic}). We focused on calculating
energy eigenstates and transition dipole moments, and studied how a
gradual symmetry lowering affects the optical selection rules.
For describing the polarizations of light associated with atomic
electric dipole oscillations we use the Jones-vector formulation, which
is introduced in Supplementary Information Section~\ref{Sec:SOM_Jones}).


\section{The resonance lines of the hydrogen atom without spin}\label{Sec:noSpin}
\subsection{The hydrogen atom in the absence of a magnetic field}

The resonance lines of electronic transitions between the $1s$ and $2p$ levels of the hydrogen atom occur around a wavelength of 120 nm.
We use the notation $|1s \rangle$ ($n = 1$; $l = 0$) and $|2p \rangle$ ($n = 2$; $l = 1$), for the ground and excited states respectively, as adopted from the book of Cohen-Tannoudji, Diu and Lalo\"{e}\cite{cohen1991quantum}. At zero magnetic field, the Hamiltonian $H_0$ (containing the kinetic and electrostatic interaction energy) of the hydrogen atom has energy eigenvalues $E_n = -E_I/n^2$, with $E_I$ the ionization energy.
Note that we will often omit the quantum number $n$, i.e.~$|s\rangle = |1s\rangle$ and $|p\rangle = |2p\rangle$.
Without considering spin, the hydrogen atom has a single $1s$ level and three degenerate $2p$ levels.
Because of this degeneracy, a single resonance line occurs, and any linear combination of orthonormal $|p \rangle$ eigenstates is a suitable eigenbasis for the Hamiltonian. A possible choice would be the basis $\{|s\rangle, |p_x\rangle, |p_y\rangle, |p_z \rangle\}$. The three p-orbitals are real-valued and have the same double-lobed shape, but are aligned along the x-, y-, and z-axes, respectively\cite{atkins2011molecular}.

A convenient measure for the strength of a transition is the real-valued oscillator strength $f$ (Supplementary Information Eq.~\ref{eq:fGen}), which is proportional to the absolute square of the transition dipole moment (which is a vector, with Cartesian components defined in Supplementary Information Eq.~\ref{eq:TDM2}). In general, the total oscillator strength $f_{tot}$ is dimensionless and for all possible transitions it adds up to the number of electrons (known as the Kuhn–Thomas sum rule\cite{atkins2011molecular}). Since we consider only a small subset of all transitions within the hydrogen atom (having $f_{tot}=1$), the total oscillator strength of our subset will be smaller than 1. However, we will consider relative values $f_{rel}$, for which the sum ($f_{rel,tot}$) will exceed 1 (see below).

The corresponding matrix elements of the transition dipole moment between the $|s\rangle$ and $|p_i\rangle$ states, with  $i \in \{ x,y,z\}$, are\cite{cohen1991quantum}

\begin{equation}\label{eq:Eq1}
\langle p_i | D_i | s \rangle =  \frac{e I_R}{\sqrt{3}}
\end{equation}
where $D_i$ is the $i$-component of $\textbf{D} = e \textbf{R}$, with $e$ the elementary charge and $\textbf{R}$ the position operator, and the constant value $I_R$ is a radial integral independent of $i$.

For an electron in the $1s$ orbital
there are three possible transitions to a $2p$ sublevel, each with a single nonzero transition dipole moment (Eq.~\ref{eq:Eq1}) and relative oscillator strength $f_{rel} = 3$ (according to Supplementary Information Eq.~\ref{eq:fRel}). The corresponding total relative oscillator strength ($3 \times 3=9$) follows from Supplementary Information Eq.~\ref{eq:fTot}, which for this case can be simplified to $f_{rel,1s,tot} = \frac{9}{(eI_R)^2} \sum\limits_{i=x,y,z}|\langle p_i | D_i | s \rangle|^2 = 9$.
Note that when only the $1s$ orbital is occupied, the absorption strength is equal for all normalized linear (complex) combinations of $D_x$, $D_y$ and $D_z$ (due to the degeneracy of the $2p$ sublevels), i.e.~the probability of a transition to $2p$ does not depend on the polarization.

\subsection{The hydrogen atom in the presence of a magnetic field}

In the presence of a static magnetic field $\textbf{B}$ along $z$, the
resonance line of the hydrogen atom is modified. The field does not only change the resonance frequencies, but also the polarization of the atomic lines, which is called the Zeeman effect.
The Hamiltonian is given by $H = H_0 + H_1$, with $H_1$ the paramagnetic coupling term (here only acting on the orbital, since we still neglect spin).
Now, the eigenbasis in which $H$ is diagonal is $\{|s \rangle, |p_{-1}\rangle, |p_0\rangle, |p_1\rangle \}$, where
\begin{equation}
\begin{aligned}
| p_{-1} \rangle &= \frac{|p_x  \rangle-i |p_y \rangle}{\sqrt{2}} \\
| p_0  \rangle &=  | p_z  \rangle \\
| p_{1} \rangle &= - \frac{|p_x  \rangle+i |p_y  \rangle}{\sqrt{2}}
\end{aligned}
\end{equation}
with the indices $-1, 0, 1$ corresponding to the $m_l$ quantum number along $z$.

\subsection{Electric dipole radiation}

In Supplementary Information Section~\ref{Sec:SOM_Dip} we perform a classical calculation of the oscillation of an electric dipole for a superposition of the ground state $| s \rangle$ and an excited state $| p_{m_l} \rangle$ of the hydrogen atom, again following \cite{cohen1991quantum}. Assuming a sample with a large number of hydrogen atoms, the result of a classical calculation equals the mean quantum mechanical value $\langle \textbf{D} \rangle_{m_l} (t)$.

For all three cases ($m_l = -1,0,1$) the mean value of the electric dipole oscillates as a function of
time, corresponding to the emission of electromagnetic energy. The type of electric dipole oscillation determines the type of polarization of the emitted radiation. Still, the polarization of light that an observer sees depends on its orientation with respect to the source (see Supplementary Information Section~\ref{Sec:SOM_Dip}).

For convenience, we will name a polarization after (the complex linear combination of) the components of $\textbf{D}$ for which (the absolute value of) the transition dipole moment is maximized. A convenient way to find this complex linear combination is the application of the Jones-vector formalism (Supplementary Information Sec.~\ref{Sec:SOM_Jones}).
For the hydrogen atom in the presence of a magnetic field (without considering spin), the matrix elements are maximized when we take
the operators $\sigma^+ = \frac{x + iy}{\sqrt{2}}$ (right circular), $\sigma^- = \frac{x - iy}{\sqrt{2}}$ (left circular) and $\pi_{z} = z$ (linear along $z$), respectively.
As such, the only non-zero matrix elements related to transitions between $1s$ and $2p$ levels are

\begin{equation}
\begin{aligned}
\langle p_{-1} | \frac{D_x - iD_y}{\sqrt{2}} |s \rangle &=  \frac{e I_R}{\sqrt{3}} \\
\langle p_0 | D_z |s \rangle &=  \frac{e I_R}{\sqrt{3}} \\
\langle p_1 | \frac{D_x + iD_y}{\sqrt{2}} | s \rangle &=  -\frac{e I_R}{\sqrt{3}}
\end{aligned}
\end{equation}
Hence, the polarization of the radiation is $\sigma^+$, $\sigma^-$ or $\pi_z$, depending on whether the non-zero matrix element is that of $\frac{D_x + iD_y}{\sqrt{2}}$, $\frac{D_x - iD_y}{\sqrt{2}}$ or $D_z$, respectively.

\begin{figure}[h!]
	\centering
	\includegraphics[width=1\textwidth]{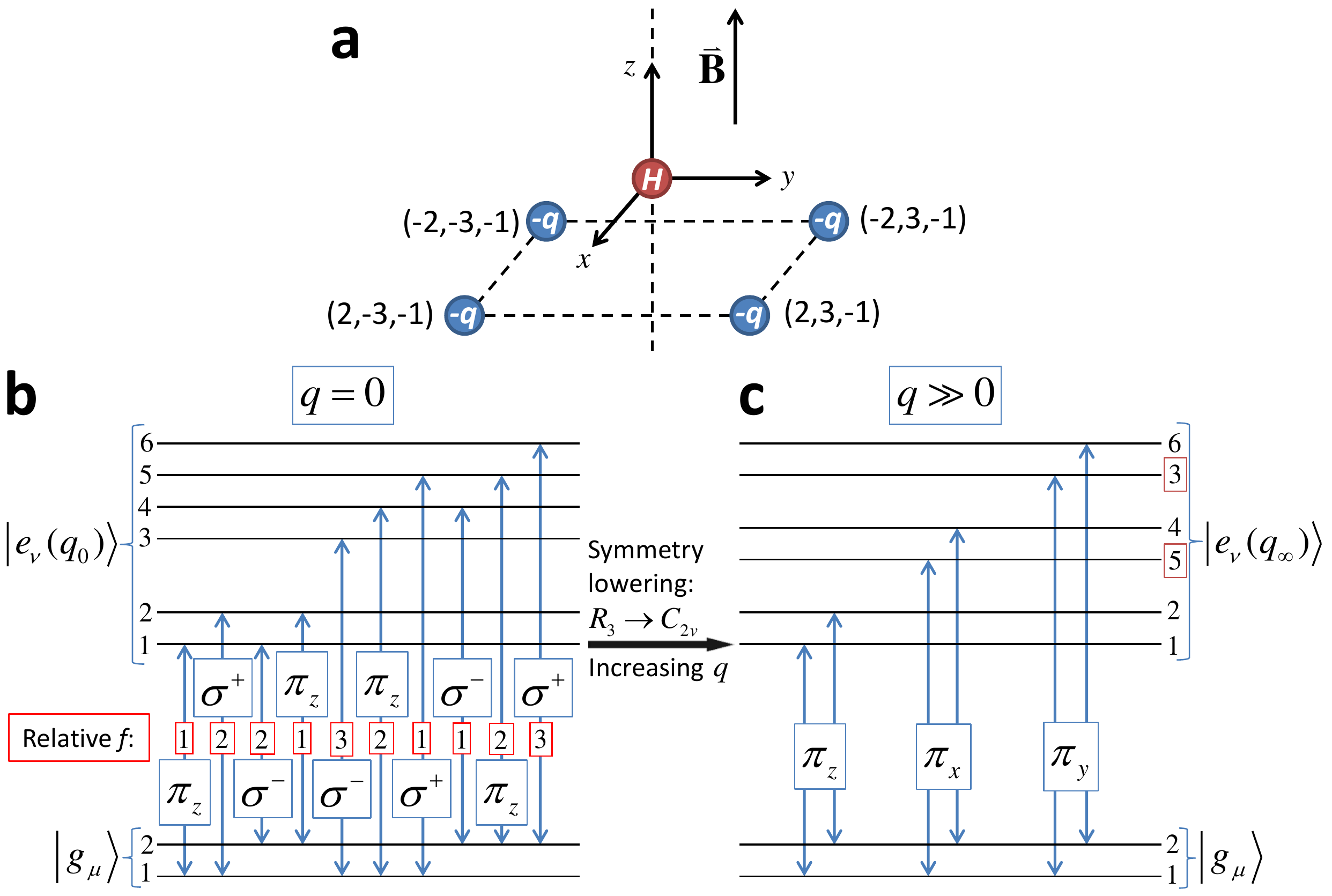}
	\caption{\textbf{Evolution of the polarization selection rules for the hydrogen atom upon symmetry lowering due to a $C_{2v}$ arrangement of negative point charges, in the presence of a weak magnetic field.}
		\textbf{a}, A hydrogen atom (in red, positioned at the origin) in a $C_{2v}$ arrangement ($C_2$ rotation axis along $z$ and two vertical mirror planes) of four negative point charges (in blue, positioned at (2,3,-1), (2,-3,-1), (-2,3,-1), (-2,-3,-1) in Bohrs). Each point charge has magnitude $-q$, where $q$ is gradually varied from $10^{-6}$ to $10^{-2}$ in atomic units. This is analogous to bringing the point charges gradually closer to the hydrogen atom.
		The weak magnetic field points in the $z$-direction.
		\textbf{b}, Energy levels with Zeeman splitting for the ground (g) state $1s$ ($g_\mu$) and excited (e) state $2p$ ($e_\nu$) sublevels of the hydrogen atom in the presence of a weak magnetic field and absence of point charges ($q_0$).
		Supplementary Information Sec.~\ref{Sec:SOM_Spin} gives a detailed analysis of the polarizations and the relative oscillator strengths $f_{rel}$ (given in red). Spin-orbit coupling (SOC) has been included.
		\textbf{c},
		In the limit of very strong	charge ($q_{\infty}$, i.e.~the perturbation due to the charges is much larger than that of the magnetic field), the excited states converge to one of the basis functions of the set $\{ |p_{z}\beta\rangle, |p_{z}\alpha\rangle, |p_{x}\beta\rangle, |p_{x}\alpha\rangle, |p_{y}\beta\rangle, |p_{y}\alpha\rangle \}$ and arrange in three doublets (split by the magnetic field). Only six transitions between the $1s$ and $2p$ levels remain allowed (equal $f$), and their polarizations are linear. Note the different ordering of $\nu$ for the excited states affected by charge.}
	\label{fig:Transitions}
\end{figure}

\section{The resonance lines of the hydrogen atom including spin}\label{Sec:withSpin}
Due to the electron and proton spins, the resonance lines of the hydrogen atom are also affected by the fine- and hyperfine structure.
In this work we will only consider the electron spin, which can be either up ($\expval{S_z}=\hbar/2$) or down ($\expval{S_z}=-\hbar/2$), to which we will refer as $\alpha$ and $\beta$, respectively. Hence, the orbitals $1s$, $2p_{-1}$, $2p_{0}$ and $2p_{1}$ allow for eight possible spinorbitals\cite{atkins2011molecular}, which are products of an orbital and a spin function. These spinorbitals form the basis $\{|s\beta \rangle, |s \alpha \rangle, |p_{-1}\beta\rangle, |p_{-1}\alpha\rangle,| p_{0}\beta\rangle, |p_{0}\alpha\rangle, |p_{1}\beta\rangle, |p_{1}\alpha\rangle \}$ to which we refer as the uncoupled representation\cite{cohen1991quantum}. Usually, these basis functions are labeled with the quantum numbers $l$, $s$, $m_l$ and $m_s$, as tabulated in Supplementary Information Table~\ref{table:266}.

The Hamiltonian $H_0$ (containing the kinetic and electrostatic interaction energy) is diagonal in this basis and the eigenvalues on the diagonal resemble the 2- and 6-fold degeneracies in energy.
When the spin-orbit coupling (SOC) term $H_{SO} = \textbf{L}\cdot\textbf{S}$ is added to the Hamiltonian, the 6-fold degeneracy of the $2p$ levels is lifted into sublevels with quantum number $j=1/2$ and $j=3/2$, i.e.~$2p_{1/2}$ (2-fold degenerate) and $2p_{3/2}$ (4-fold)\cite{cohen1991quantum}. Here, $j$ is the total angular momentum quantum number related to $J^2$ (with eigenvalues $\hbar^2 j(j+1)$), whereas the total angular momentum is $\textbf{J} = \textbf{L} + \textbf{S}$ (Supplementary Information Section~\ref{Sec:SOM_SOC}).
The degeneracy can be further lifted by e.g.~a magnetic field (which introduces an additional term to $H$).
Also, the magnetic field induces a quantization axis. It is now convenient to use an approach based on time-independent degenerate perturbation theory\cite{cohen1991quantum}.
Since the field is applied in the $z$-direction,
we define a basis
formed by the eigenstates of the total angular momentum $J_z$. Constructing the matrix $J_z = L_z + S_z$ in the basis in which $H$ is diagonal, one finds that $J_z$ is block-diagonal, i.e. the basis does not necessarily consist of eigenstates of $J_z$. We determine the eigenfunctions of $J_z$ via diagonalization of the 2- and 4-fold degenerate subspaces, which provide the basis to which we refer as the coupled representation
(where $H$ remains diagonal), which is the convenient one for the case of SOC.

Good quantum numbers are now $j$, $m_j$, $l$ and $s$. The basis can be expressed as a linear combination of the basis functions of the uncoupled representation (Supplementary Information Table~\ref{table:277}), where the prefactors are the so-called Clebsch-Gordan coefficients\cite{cohen1991quantum}.

A transition (via excitation or emission) between a $1s$ and $2p$ sublevel is possible if a nonzero value is obtained for the transition dipole moment
$\langle p_{j,m_j} | \textbf{D} |s_{m_j} \rangle$, with
$|s_{m_j} \rangle = |j=\frac{1}{2},m_j = \pm \frac{1}{2},l=0,s=\frac{1}{2} \rangle$ and
$|p_{j,m_j} \rangle = |j,m_j,l=1,s=\frac{1}{2} \rangle$.
The corresponding matrix elements have been calculated
in Supplementary Information Table~\ref{table:288}.
The relative oscillator strength $f_{rel}$ is given by Supplementary Information Eq.~\ref{eq:fRel}, which directly depends on the Clebsch-Gordan coefficients.
We see that now ten of the twelve
possible transitions between a $1s$ and $2p$ sublevel
have nonzero oscillator strength. In contrast, only six
transitions are allowed when SOC is not taken into account (three for either up or down spin).
We will determine the polarizations and $f_{rel}$-values also numerically in Section~\ref{Sec:C2v}, where the $1s$ and $2p$ sublevels will be denoted as $|g_\mu \rangle$ and $|e_\nu \rangle$, respectively (Fig.~\ref{fig:Transitions}b).

\section{Evolution of optical selection rules for the hydrogen atom in a $C_{2v}$ arrangement of point charges}\label{Sec:C2v}

In this section, we report on ab initio calculations that study the evolution of the optical selection rules
for transitions between the 1s and 2p sublevels of the hydrogen atom upon gradual symmetry lowering due to a $C_{2v}$ arrangement of negative point charges (each with magnitude $-q$), in the presence of a weak magnetic field (Fig.~\ref{fig:Transitions}a). Such a relatively simple system is already too complicated to solve in an analytical way, such that we have to use numerical
methods.
First, we numerically calculate functions that are relatively good approximations for the eigenstates of the Hamiltonian. Strictly speaking, these functions are not eigenstates because one has to take a finite basis set. Nevertheless, we will often refer to these functions as states (or eigenfunctions). Secondly, we numerically calculate the
Cartesian components of the corresponding transition dipole moments (with the relevant ones defined in Supplementary Information Eq.~\ref{eq:TDM2}). An accurate way to calculate these is the use of the CASSCF/RASSI\textendash SO method (which combines the Complete Active Space Self Consistent Field (CASSCF)
and Restricted Active Space State Interaction (RASSI) method with the
inclusion of SOC),
as introduced by Roos and Malmqvist\cite{roos2004effects,roos2004relativistic}. We perform
such calculations using the MOLCAS\cite{anderssonmolcas} software. To approximate the $1s$ and $2p$-orbital we use the large ANO basis set\cite{widmark1990density}, which for the excited states of the hydrogen atom does actually not very accurately approximate the energies. Our purpose however is to illustrate how the mixing of sublevels affects transition dipole moments. In this regard, the quality of the orbitals is expected to be sufficient, since they have the required symmetry.

Including a magnetic field within ab initio calculations is not straightforward. We will therefore mimic the field by inducing a quantization axis $z$ (Fig.~\ref{fig:Transitions}a), through diagonalization of $J_z$ within degenerate subspaces (see Section~\ref{Sec:withSpin}). This provides the required eigenbasis to which the calculated transition dipole moments are transformed.
From the transition dipole moments, we can calculate the relative oscillator strength $f_{rel}$ for each
transition between a $1s$ and $2p$ sublevel, according to Supplementary Information Eq.~\ref{eq:fRel}. The evolution of $f_{rel}$ as a function of $q$ is depicted in Fig.~\ref{fig:OscStr}.

\begin{figure}[h!]
	\centering
	\includegraphics[width=5cm]{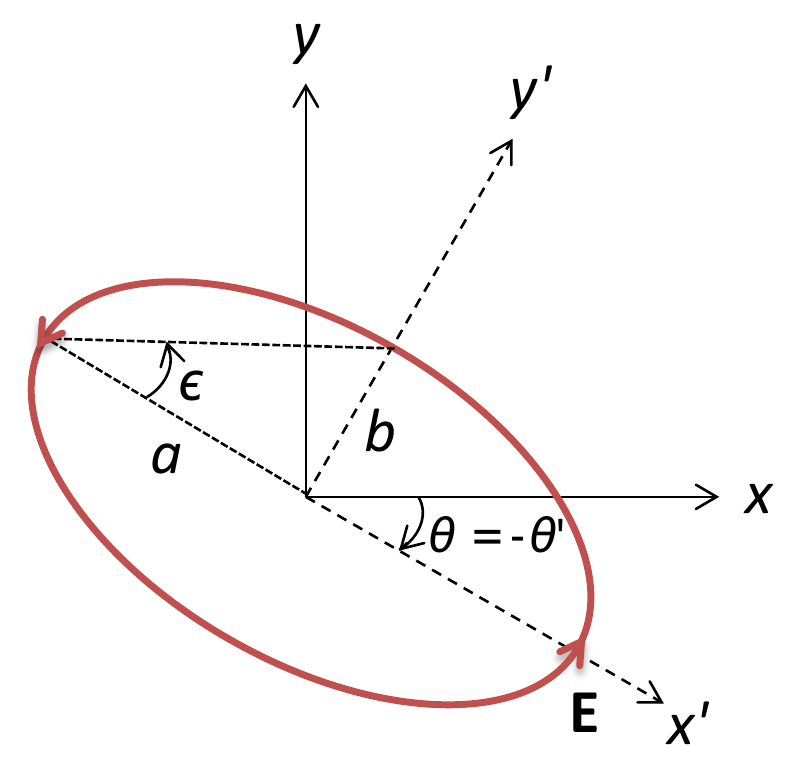}
	\caption{\textbf{The polarization ellipse.} The main parameters that define the polarization ellipse\cite{r1987ellipsometry} are the azimuth $\theta$ ($-\frac{1}{2} \pi \leqslant \theta < \frac{1}{2} \pi$) of the semi-major axis $a$ with respect to the $x$-axis, and the ellipticity angle $\epsilon$ ($-\frac{1}{4} \pi \leqslant \epsilon \leqslant \frac{1}{4} \pi$), which is defined through the ellipticity $e = \frac{b}{a}$ (with $b$ the semi-minor axis) such that $e = \pm \tan \epsilon$, where the $+$ and $-$ signs correspond to right- and left-handed polarization respectively. In the figure the indicated polarization is left-handed, i.e.~the electric vector
		oscillates counterclockwise. The total amplitude of the electric field is given by $A = \sqrt{a^2+b^2}$, where one usually takes an amplitude of $A=1$.
		It is also common to take a global phase factor $\delta = 0$.
	}
	\label{fig:PolarizationEllipse}
\end{figure}

We will use the Jones-vector formulation (see also Supplementary Information Section~\ref{Sec:SOM_Jones}) to
investigate how the polarization selection rules are affected as a function of $q$.
The Jones-vector formulation assigns a polarization ellipse (Fig.~\ref{fig:PolarizationEllipse}) with azimuth $\theta$ ($-\frac{1}{2} \pi \leqslant \theta < \frac{1}{2} \pi$) and ellipticity angle $\epsilon$ ($-\frac{1}{4} \pi \leqslant \epsilon \leqslant \frac{1}{4} \pi$) to the oscillation of an electric vector\cite{r1987ellipsometry}. Normally, this electric vector is the electric field component of a light wave. Instead, we will assign such a polarization ellipse to the oscillation of an atomic electric dipole related to an electronic transition, with the components of the electric vector given by the (normalized) components of the corresponding transition dipole moment (Supplementary Information Eq.~\ref{eq:Evector}).
A convenient method to visualize the Jones vector is via the Poincar\'{e}-sphere representation\cite{r1987ellipsometry}.
Within this method, the longitude $2\theta$ and latitude $2\epsilon$ determine a point (labeled $P$ in Fig.~\ref{fig:Ellipticity}a) representing the ellipse of polarization with azimuth $\theta$ and ellipticity angle $\epsilon$ (Fig.~\ref{fig:PolarizationEllipse}).

\begin{figure}[h!]
	\centering
	\includegraphics[width=1.0\textwidth]{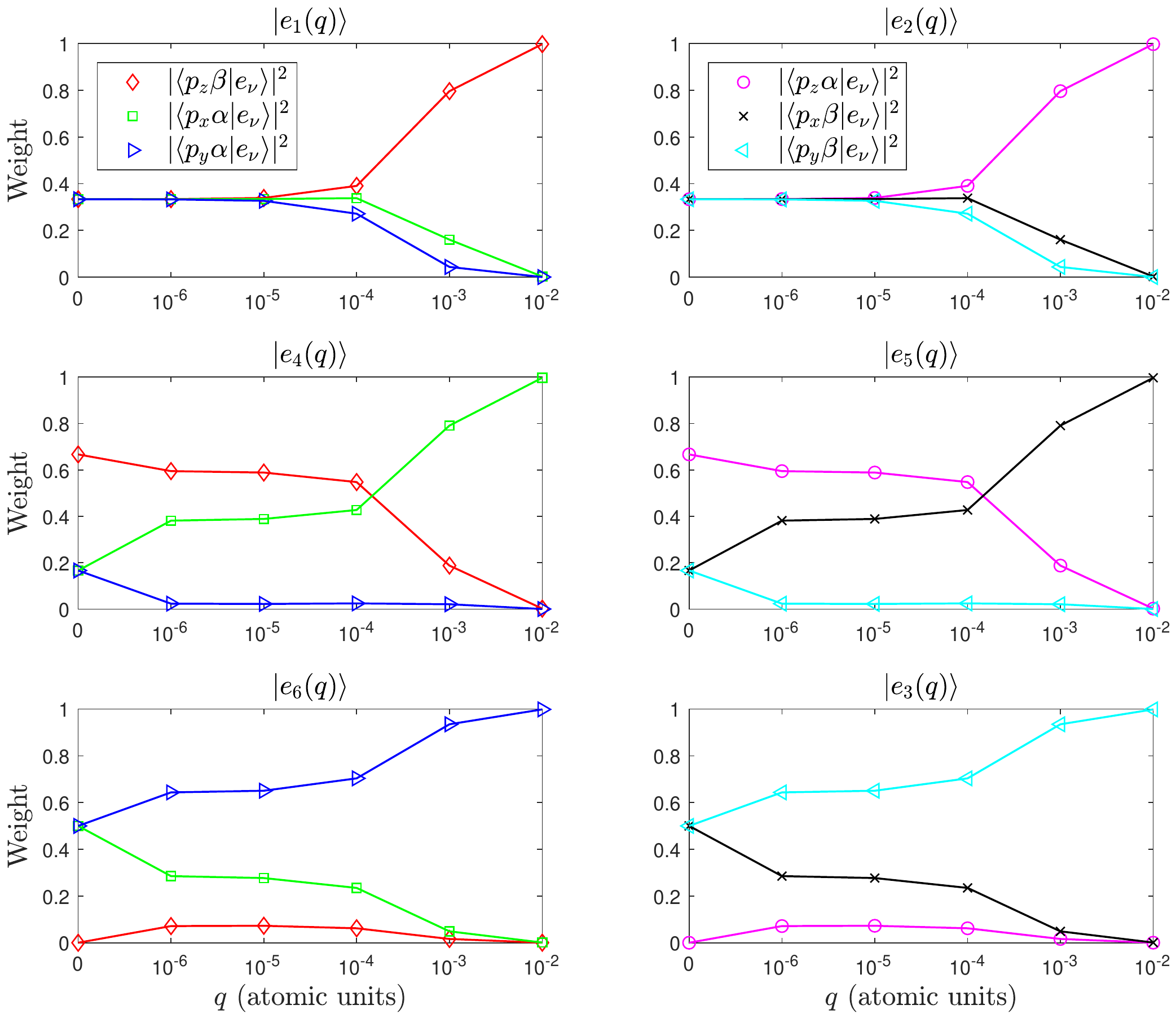}
	\caption{\textbf{Evolution of the excited states for the hydrogen atom upon gradual symmetry lowering due to a $C_{2v}$ arrangement of negative point charges, in the presence of a weak magnetic field.}
		Weights
		(absolute squares of the coefficients)
		for the
		excited states $|e_\nu(q)\rangle$ with $-q$ the magnitude of the point charges in atomic units and $\nu \in \{1,...,6\}$
		as written in the basis as used for the bare H atom, i.e.~$\{|s \beta \rangle, |s \alpha\rangle, |p_{z}\beta\rangle, |p_{z}\alpha\rangle, |p_{x}\beta\rangle, |p_{x}\alpha\rangle, |p_{y}\beta\rangle, |p_{y}\alpha\rangle \}$. The left plots ($\nu \in \{1,4,6\}$) have in common that $|e_\nu(q)\rangle$ is a superposition of the states $|p_{z}\beta\rangle$, $|p_{x}\alpha\rangle$ and $|p_{y}\alpha\rangle$, whereas for the right plots $|e_\nu(q)\rangle$ is a superposition of $|p_{z}\alpha\rangle$, $|p_{x}\beta\rangle$ and $|p_{y}\beta\rangle$. Furthermore, the plots are ordered in rows based on the fact that the weights as a function of $q$ are the same within each row.
		In the limit of very strong
		charge ($q_{\infty}$, i.e.~the perturbation due to the charges is much larger than due to the magnetic field), the excited states converge to one of the basis states of the set $|\{p_{z}\beta\rangle, |p_{z}\alpha\rangle, |p_{x}\beta\rangle, |p_{x}\alpha\rangle, |p_{y}\beta\rangle, |p_{y}\alpha\rangle \}$.
		For the first row, there is convergence towards $|p_z\rangle$, towards $|p_x\rangle$ for the second, and $|p_y\rangle$ for the third.
		Data points are connected to guide the eye.}
	\label{fig:S3toS8}
\end{figure}

\subsection{The hydrogen atom in the presence of a weak magnetic field}

As a proof of principle calculation, we will first perform numerical calculations on the pure atom (i.e.~$q=q_0=0$), to see whether we obtain the same optical selection rules as the analytical solution.
Fig.~\ref{fig:Transitions}b considers a hydrogen atom in the presence of a magnetic field in the $z$-direction (without point charges), where
we label the $1s$ and $2p$ sublevels as the ground and excited state sublevels $|g_{\mu} \rangle$ ($\mu \in \{1,2\}$) and $|e_{\nu}(q=q_0) \rangle$ ($\nu \in \{1,...,6\}$), respectively. As the notation indicates (where the dependence on $q$ is already introduced for later use), only the excited states depend on the charge $q$. For the case of $q_0$, the numbering of $\mu$ and $\nu$ increases with increasing energy
for both ground and excited states.

From the CASSCF/RASSI\textendash SO calculations, the eigenfunctions of the Hamiltonian are obtained. To mimic the magnetic field in the $z$-direction, a quantization axis $z$ is induced by diagonalization of $J_z$ within the 2- and 4-fold degenerate subspaces.
As such, we obtain the states $|g_{\mu} \rangle$
and $|e_{\nu} (q_0) \rangle$ (Table~\ref{table:conv}), which are the same states (apart from a global phase factor) as those obtained from the analytical solution, i.e.~the coupled representation (Supplementary Information Table~\ref{table:277}). The weights (i.e.~the absolute squares of the coefficients when decomposing as in Eq.~\ref{eq:dec2} and~\ref{eq:dec2}) of the excited states have been plotted in Fig.~\ref{fig:S3toS8} (first data point of each subplot corresponds to $q=0$).

\begin{figure}[h!]
	\centering
	\includegraphics[width=1.0\textwidth]{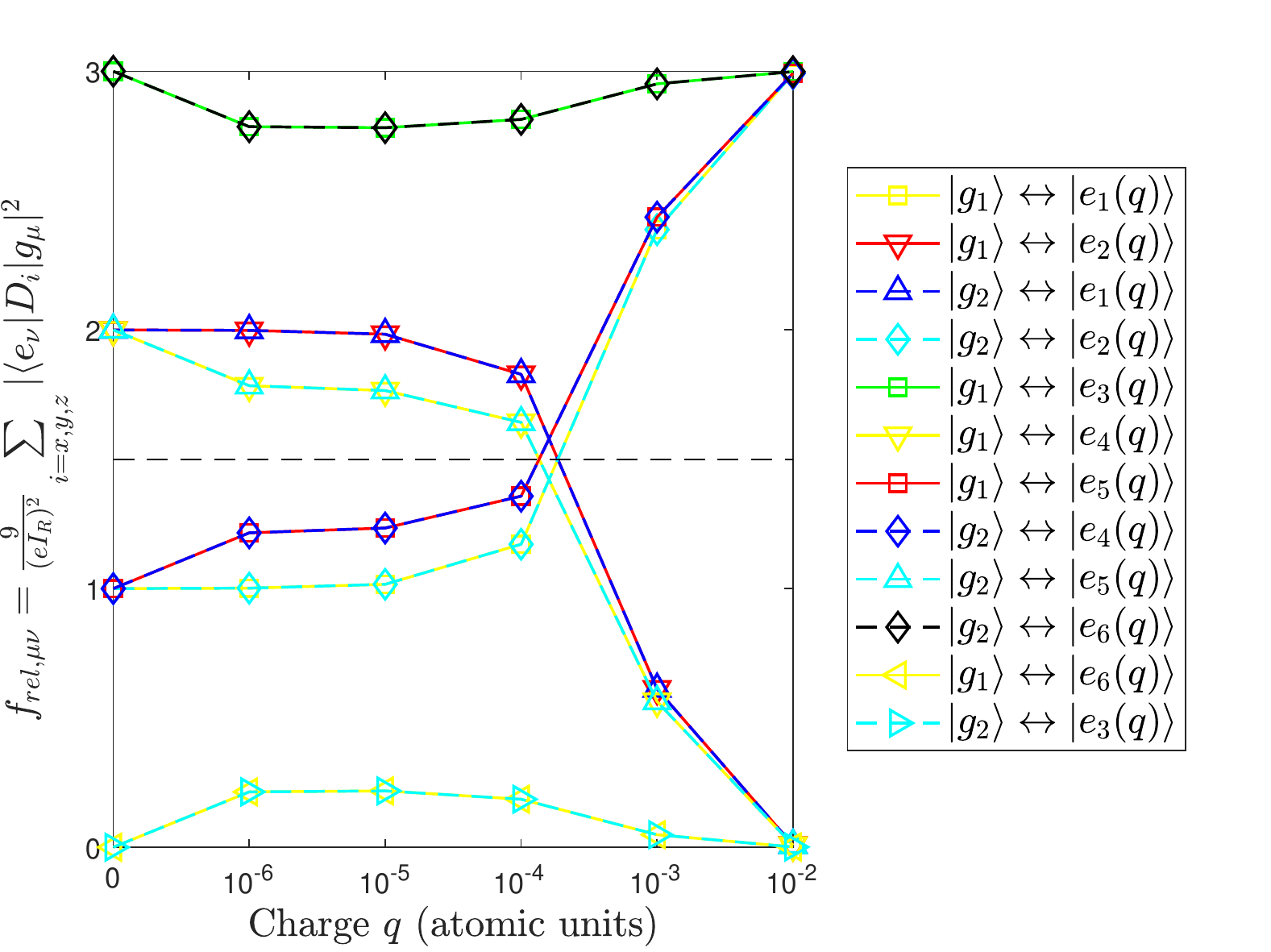}
	\caption{\textbf{Evolution of the relative oscillator strengths for transitions between $1s$ ($g_{\mu}$) and $2p$ ($e_{\nu}$) sublevels of the hydrogen atom upon gradual symmetry lowering due to a $C_{2v}$ arrangement of negative point charges, in the presence of a weak magnetic field.}
		The relative oscillator strength $f_{rel,\mu\nu}$ for a transition between $| g_{\mu} \rangle$ and $| e_{\nu} \rangle$ has been defined in Supplementary Information Eq.~\ref{eq:fRel}.
		Interestingly, the two originally forbidden transitions become slightly allowed ($\pi_z$ polarization) for small $q$-values (forbidden for $q_0$ and $q_{\infty}$). Note that the sum of the relative $f$-values does not vary as a function of $q$, i.e.~$f_{rel,1s\alpha,tot} = f_{rel,1s\beta,tot} = 9$. This becomes particularly clear from the fact that the plot has a horizontal mirror plane (dashed line) at $f=1.5$.
		Data points are connected to guide the eye. See Table~\ref{table:pola} for the $q$-dependent optical selection rules.}
	\label{fig:OscStr}
\end{figure}

The CASSCF/RASSI\textendash SO calculations also provide transition dipole moments.
We transform the matrix elements to the basis obtained after diagonalization of $J_z$ within the degenerate subspaces. Now we have obtained the $i$-components $\langle e_\nu(q_0) | D_i |g_\mu \rangle$ ($i \in \{ x,y,z\}$) of the transition dipole moment related to the $|g_\mu\rangle \leftrightarrow |e_\nu(q_0) \rangle$ transitions.
From Supplementary Information Eq.~\ref{eq:fRel}, the corresponding relative oscillator strengths ($f_{rel}$) are obtained, which are the first data points ($q=0$) of each series in Fig.~\ref{fig:OscStr}. Since the numerical $f_{rel}$-values are exactly the same as the analytical ones (Supplementary Information Table~\ref{table:288}), i.e.~$f_{rel}\in\{0,1,2,3\}$, we conclude that our method is accurate.

Using the Jones-vector formalism (see also Supplementary Information Section~\ref{Sec:SOM_Jones}), our numerical calculations also provide the same polarization selection rules as in Supplementary Information Table~\ref{table:288}. The evolution of the optical selection rules as a function of $q$ for the six transitions having their electric dipole oscillating in the $xy$-plane has been visualized in Fig.~\ref{fig:Ellipticity}a and~b, where the first data point of each series corresponds to $q=0$, for which the transitions are circular.

\subsection{The hydrogen atom in the presence of a $C_{2v}$ arrangement of point charges and a weak magnetic field}

To study the dependence of the atomic electric dipole oscillation on a charged environment, we consider the hydrogen atom in a $C_{2v}$ arrangement of four negative point charges $-q$ (positioned
at (2,3,-1), (2,-3,-1), (-2,3,-1), (-2,-3,-1) in Bohrs with respect to the hydrogen atom), as depicted in Fig.~\ref{fig:Transitions}a.
The symmetry of the hydrogen atom is gradually distorted by increasing $q$. This is analogous to studying the effect of charges gradually approaching a hydrogen atom, starting at infinite distance.
Such a gradual symmetry distortion will gradually affect the Hamiltonian and its eigenstates.

Again, from the CASSCF/RASSI\textendash SO calculations, the eigenfunctions of the Hamiltonian and
the transition dipole moments are obtained. In the absence of a magnetic field and in the presence of charges, the six excited states form three doublets.
A magnetic field will further lift these degeneracies and impose additional optical selection rules. To mimic a magnetic field in the $z$-direction, a quantization axis $z$ is induced by diagonalization of $J_z$ within the 2-fold degenerate subspaces.
As such, the states $|g_{\mu} \rangle$ (for which there is no dependence on $q$) and $|e_{\nu}(q\ne 0) \rangle$ are obtained, where we assume that the perturbation due to the magnetic field is much weaker than due to the charges (i.e.~the magnetic field affects the energies only by slightly lifting the degeneracies).

The excited states depend on the magnitude of the surrounding charges and are denoted as $|e_{\nu}(q) \rangle $, with $\nu \in \{ 1,...,6 \}$.
The labeling of $\nu$ is based on the evolution of these coefficients: with each gradual increase of charge ($q \rightarrow q'$), the states are slightly affected and the new state $|e_{\nu'}(q') \rangle$ is labeled with the $\nu$-value that most resembles $|e_\nu(q)\rangle$ (i.e.~$\nu'=\nu$ for the $\nu'$ with largest overlap $\langle e_\nu(q) |e_{\nu'}(q')\rangle$). Accordingly, we find a different order of $\nu$ for the excited states affected by charge (Fig.~\ref{fig:Transitions}c).
Whereas $q_0$ (Fig.~\ref{fig:Transitions}b) denotes the absence of charge, $q_{\infty}$ (Fig.~\ref{fig:Transitions}c) denotes the case of very strong charge (but small enough to not ionize the hydrogen atom).

It turns out that we can write the excited states always as a linear combination of at most three basis functions of the set $\{|p_{z}\beta\rangle, |p_{z}\alpha\rangle,| p_{x}\beta\rangle, |p_{x}\alpha\rangle, |p_{y}\beta\rangle, |p_{y}\alpha\rangle \}$. We find the following relations (where the coefficients are the projections onto each of the basis functions)
\begin{equation} \label{eq:dec1}
| e_{\nu=1,4,6}(q) \rangle = \langle p_z \beta | e_\nu(q) \rangle |p_z \beta \rangle + \langle p_x \alpha | e_\nu(q) \rangle |p_x \alpha \rangle + \langle p_y \alpha | e_\nu(q) \rangle |p_y \alpha \rangle
\end{equation}
\begin{equation} \label{eq:dec2}
| e_{\nu=2,5,3} (q) \rangle = \langle p_z \alpha | e_\nu(q) \rangle |p_z \alpha \rangle + \langle p_x \beta | e_\nu(q) \rangle |p_x \beta \rangle + \langle p_y \beta | e_\nu(q) \rangle |p_y \beta \rangle
\end{equation}
of which the weights (absolute squares of the coefficients) are presented in Fig.~\ref{fig:S3toS8}.
For each of the doublets (first row in Fig.~\ref{fig:S3toS8}: $\nu=1,2$; second row: $\nu=4,5$; third row: $\nu=6,3$) the weights of the two different sublevels are the same.

In the limit of very strong
charge ($q_{\infty}$), the excited states converge to one of the basis functions of the set $\{|p_{z}\beta\rangle, |p_{z}\alpha\rangle,| p_{x}\beta\rangle, |p_{x}\alpha\rangle, |p_{y}\beta\rangle, |p_{y}\alpha\rangle \}$ (see Fig.~\ref{fig:S3toS8} and Table~\ref{table:conv}).
Consequently, a nonzero value for the transition dipole moment is only obtained for transitions between sublevels with equal spin, e.g.~$\langle p_x \alpha | D_x |s \alpha \rangle = \langle p_x | D_x |s \rangle \langle \alpha | \alpha \rangle = \frac{e I_R}{\sqrt{3}}$ (see also Eq.~\ref{eq:Eq1}). As such, only six of the twelve transitions are allowed (ten for $q_0$), with linear polarization $\pi_x$, $\pi_y$ or $\pi_z$, depending on whether the excited state is $| p_x \rangle$, $|p_y\rangle$ or $|p_z\rangle$, respectively (see  Fig.~\ref{fig:Transitions}c). Apparently, the interaction with the surrounding charges outweighs the contribution from SOC, such that only spin-conserving transitions are allowed.

\begin{table}[t]
	\centering
	\caption{\textbf{The excited states of the hydrogen atom upon symmetry lowering
			due to a $C_{2v}$ arrangement of negative point charges.}
		The excited states $| e_{\nu}(q) \rangle$ are tabulated for the case of zero charge ($q_{0}$) and very strong charge ($q_{\infty}$). In the latter case the final excited states converge to one of the basis functions of the set $\{|p_{z}\beta\rangle, |p_{z}\alpha\rangle,| p_{x}\beta\rangle, |p_{x}\alpha\rangle, |p_{y}\beta\rangle, |p_{y}\alpha\rangle \}$. The ground states $|g_{\mu} \rangle$ ($|s \beta \rangle$ and $|s \alpha \rangle$) remain unaffected as a function of $q$.}
	\begin{tabular}{|c | c| c|}
		\hline
		$\nu$ & $| e_\nu(q_0) \rangle$ &	$| e_\nu(q_{\infty}) \rangle$  \\ [0.5ex]
		\hline 		
		$1$ & $\frac{1}{\sqrt{3}}(| p_z \beta \rangle - |p_x \alpha \rangle + i |p_y \alpha \rangle)$ & $|p_z \beta \rangle$ \\
		$2$ & $\frac{1}{\sqrt{3}}(| p_z \alpha \rangle + |p_x \beta \rangle + i |p_y \beta \rangle)$ & $|p_z \alpha \rangle$ \\
		$3$ & $-\frac{1}{\sqrt{2}}(i |p_x \beta \rangle + |p_y \beta \rangle)$ & $|p_y \beta \rangle$ \\
		$4$ & $-\sqrt{\frac{2}{3}}| p_z \beta \rangle +\frac{1}{\sqrt{6}}(-|p_x \alpha \rangle + i |p_y \alpha \rangle)$ & $|p_x \alpha \rangle$ \\
		$5$ & $-\sqrt{\frac{2}{3}}| p_z \alpha \rangle +\frac{1}{\sqrt{6}}(|p_x \beta \rangle + i |p_y \beta \rangle)$ & $|p_x \beta \rangle$ \\
		$6$ & $\frac{1}{\sqrt{2}}(-i |p_x \alpha \rangle + |p_y \alpha \rangle)$ & $|p_y \alpha \rangle$ \\ [1ex]	
		\hline
	\end{tabular}
	\label{table:conv}
\end{table}

To study the optical selection rules for intermediate $q$-values, we use the transition dipole moments as obtained from our CASSCF/RASSI\textendash SO calculations. Again, we transform the transition dipole moments to the basis obtained after diagonalization of $J_z$ within the degenerate subspaces, such that we obtain the $i$-components $\langle e_\nu(q) | D_i |g_\mu \rangle$ ($i \in \{ x,y,z\}$) of the transition dipole moment related to the $|g_\mu\rangle \leftrightarrow |e_\nu(q) \rangle$ transitions.
Interestingly, we can divide the twelve possible transitions into two groups (Table~\ref{table:pola}), based on the direction in which the electric dipole oscillates. For Group XY (red) it oscillates in the $xy$-plane (the $z$-component of the transition dipole moment remains zero for increasing $q$). For Group Z (blue) it oscillates in the $z$-direction (zero $x$- and $y$-components).

\begin{table}[t]
	\centering
	\caption{\textbf{Evolution of the polarization selection rules of $|g_\mu \rangle \leftrightarrow |e_\nu (q)\rangle$
			transitions for the hydrogen atom upon symmetry lowering
			due to a $C_{2v}$ arrangement of negative point charges.}
		The twelve possible $|g_\mu \rangle \leftrightarrow |e_\nu (q)\rangle$ transitions are divided into two groups (both containing six transitions), based on the direction in which the electric dipole oscillates: for Group XY (red) it oscillates in the $xy$-plane (the $z$-component of the transition dipole moment remains zero for increasing $q$), whereas for Group Z (blue) it oscillates in the $z$-direction (zero $x$- and $y$-components). Cells with an arrow denote how the polarization changes from $q_0$ (left value) to $q_\infty$ (right), where a zero denotes a transition with zero oscillator strength $f$.
		Two cells contain only $\pi_z$, implying that the polarization is unaffected (although $f$ increases with $q$). The values $0(\pi_x)$ denote a polarization change towards $\pi_x$, whereas $ \lim\limits_{q \rightarrow \infty} f(q) = 0$. The two originally forbidden transitions become slightly allowed ($\pi_z$-polarized) for nonzero $q$, but $ \lim\limits_{q \rightarrow \infty} f(q) = 0$, which is denoted as $0(\pi_z)$.}
	\begin{tabular}{|c | c| c|c | c| c|c|}
		\hline
		\diagbox[width=1.5em]{$\mu$}{$\nu$}  & 1 &2&3&4&5&6 \\ [0.5ex]
		\hline 		
		1 & \textcolor{blue}{$\pi_z$} & \textcolor{red}{$\sigma^+ \rightarrow 0 (\pi_x)$}& \textcolor{red}{$\sigma^- \rightarrow \pi_y$}&\textcolor{blue}{$\pi_z \rightarrow 0$} &\textcolor{red}{$\sigma^+ \rightarrow \pi_x$}&\textcolor{blue}{$0 \rightarrow 0(\pi_z)$} \\
		2 &\textcolor{red}{$\sigma^- \rightarrow 0(\pi_x)$} &\textcolor{blue}{$\pi_z$}&\textcolor{blue}{$0 \rightarrow 0(\pi_z)$}&\textcolor{red}{$\sigma^- \rightarrow \pi_x$}&\textcolor{blue}{$\pi_z \rightarrow 0$}&\textcolor{red}{$\sigma^+ \rightarrow \pi_y$}	 \\ [1ex]	
		\hline
	\end{tabular}
	\label{table:pola}
\end{table}

Fig.~\ref{fig:OscStr} depicts the gradual evolution of the relative oscillator strength ($f_{rel}$) as a function of $q$ for all twelve $|g_\mu\rangle \leftrightarrow |e_\nu (q)\rangle$ transitions, as obtained from Supplementary Information Eq.~\ref{eq:fRel}.
Although certain transitions become even forbidden with increasing charge, total emission and absorption remain the same, because the sum of the oscillator strengths of all transitions from or to $|1s\alpha \rangle $ or $|1s\beta \rangle$ is unaffected, i.e.~$f_{rel,1s\alpha,tot} = f_{rel,1s\beta,tot} = 9$ (compare Section~\ref{Sec:noSpin} for the orbitals and $f_{rel}$-values for the $q_{\infty}$-case, and Supplementary Information Table~\ref{table:288} for the $f_{rel}$-values when $q=0$).
This becomes also clear from the fact that the plot has a horizontal mirror plane (dashed line) at $f=1.5$.
Since the transitions have different polarizations, the dependence of the oscillator strengths of each transition on $q$ implies that the amount of light emitted in a specific direction depends on $q$ as well. However, when light would be collected from all directions simultaneously, no variation would be observed in the intensity.
Particularly interesting is the fact that the two originally forbidden transitions ($q = 0$ and SOC included, see Supplementary Information Table~\ref{table:288}) become slightly allowed ($\pi_z$ polarization) for $0 < q < \infty $, which results from the fact that $|e_\nu(3)(q)\rangle$ and $|e_\nu(6)(q)\rangle$ gain some contribution from $|p_z \alpha \rangle$ and $|p_z \beta \rangle$, respectively (Fig.~\ref{fig:S3toS8}).

\begin{figure}[t]
	\centering
	\includegraphics[width=1\textwidth]{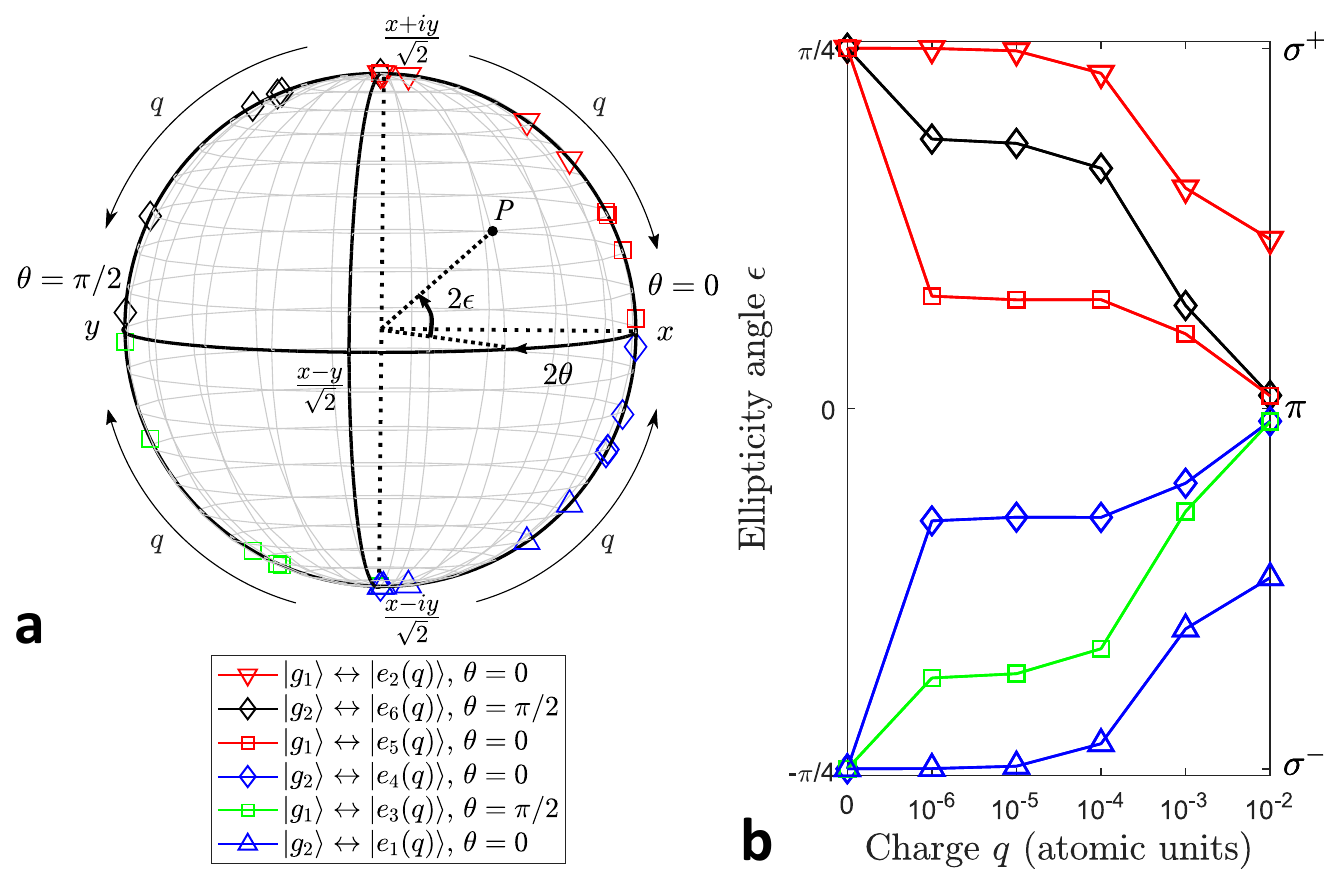}
	\caption{\textbf{Evolution of the ellipticity angle $\epsilon$ and azimuth $\theta$ for the hydrogen atom upon gradual symmetry lowering.} We consider the six transitions (see legend) that have the electric dipole oscillating in the $xy$-plane (Group XY in Table~\ref{table:pola}).
		\textbf{a}, For these six transitions, the polarization change as a function of $q$
		is represented on a Poincar\'{e} sphere\cite{r1987ellipsometry}.
		The longitude $-\pi \leqslant 2\theta < \pi$ and latitude $-\frac{1}{2} \pi \leqslant 2\epsilon \leqslant \frac{1}{2} \pi$ determine a point $P$, that represents a polarization ellipse with azimuth $\theta$ and ellipticity angle $\epsilon$ (Fig.~\ref{fig:PolarizationEllipse}). For the green (squares) and black (diamond) series the $y$-axis is the major axis, hence $\theta = \pi/2$. For the other four (red and blue) series the $x$-axis is the major axis of the polarization ellipse, hence $\theta = 0$. The four arrows outside the sphere indicate for the data points the direction of increasing $q$.
		\textbf{b}, Ellipticity angle values $\epsilon$ from ($a$) as a function of charge $q$ for the six different transitions, changing all from circular ($\epsilon = \pm \pi/4$) towards linear ($\epsilon = 0$). Data points are connected to guide the eye.}
	\label{fig:Ellipticity}
\end{figure}

To study the gradual evolution of the optical selection rules as a function of $q$, we use the Jones-vector formalism (see also Supplementary Information Section~\ref{Sec:SOM_Jones}).
For the six transitions having the electric dipole oscillating in the $xy$-plane (Group XY, red in Table~\ref{table:pola}), the Jones vectors are visualized in Fig.~\ref{fig:Ellipticity}a via the Poincar\'{e}-sphere representation\cite{r1987ellipsometry}, and the ellipticity angles $\epsilon$ are plotted as a function of $q$ in Fig.~\ref{fig:Ellipticity}b.
We find for all six Group XY transitions that the polarization of the atomic electric dipoles changes gradually upon gradual increase of the point charges. More specific, the polarization changes from circular ($\sigma$) via elliptical towards linear ($\pi$). The change to linear goes most rapid for the transitions where $|e_4 \rangle$ and $|e_5 \rangle$ are involved, a bit slower for the transitions with $|e_3 \rangle$ and $|e_6 \rangle$, and slowest for the transitions with $|e_1 \rangle$ and $|e_2 \rangle$ (which actually become forbidden for large $q$). This corresponds, respectively, to excited states that evolve towards $|p_x \rangle$, $|p_y \rangle$ and $|p_z \rangle$ character (see also Fig.~\ref{fig:S3toS8}). The fact that the evolution towards linear polarization goes faster for the transitions associated with $|p_x \rangle$ than for the ones associated with $p_y$ is related to the particular design of the distortion used in our study: the charges $q$ are in $x$-direction closer to the atom than in $y$-direction (see Fig.~\ref{fig:Transitions}a).

For the six allowed transitions at $q_{\infty}$, the $\pi_z$-transitions originate from $\pi_z$ (for $q_{0}$), whereas $\pi_x$ and $\pi_y$ originate from $\sigma$ (and are elliptical for intermediate $q$-values). An observer at the $+z$-direction will (with gradually increasing charge) see that the polarization of emitted light changes gradually from circular to linear.
Similarly, absorption of light becomes with increasing $q$ ultimately most efficient for linearly polarized light.

\section{Summary and Outlook}\label{Sec:Sum}
Studying the electronic transitions of the $1s$ and $2p$ levels of the hydrogen atom in the presence of negative point charges (and a weak magnetic field)
provided a better understanding of the relation between the electronic wavefunctions and the polarizations of the interacting light. By external lowering of the symmetry of the hydrogen atom by gradually changing the magnitude of negative point charges in a $C_{2v}$ arrangement, it was found that the polarization selection rules were affected gradually as well (both oscillator strength and polarization). Only six transitions (equal oscillator strength and linear polarization) remained allowed between $1s$ and $2p$ sublevels in the limit of very strong charge.

This study has provided a simple model system to show the principle of symmetry dependent optical selection rules.
We have shown for the hydrogen atom that varying the magnitude of negative point charges allows to switch the optical selection rules of certain transitions between circular and linear (elliptical in between) and to switch other transitions between allowed (on) and forbidden (off). Such switching could be interesting for the storage and transfer of (quantum) information.
The study also provides a better intuition for
polarization selection rules of systems with (relatively) low symmetry (like molecules or crystal defects).


\vspace{1cm}
\noindent \textbf{Acknowledgements}\\
We thank the Center for Information Technology of the University of Groningen
for their support and access to the Peregrine high performance computing cluster. Financial support was provided by the BIS grant of the Zernike Institute for Advanced Materials.


\vspace{1cm}
\noindent \textbf{Author Contributions}\\ The project was initiated by all authors. Calculations and data analysis were performed by G.J.J.L, and he had the lead on writing the manuscript. All authors contributed to improving the manuscript.


\vspace{1cm}
\noindent \textbf{REFERENCES}

\vspace{-16mm}


\clearpage

\setcounter{table}{0}
\setcounter{figure}{0}
\setcounter{section}{0}
\setcounter{page}{1}
\setcounter{tocdepth}{1}

\renewcommand{\thesection}{\arabic{section}}
\renewcommand{\thefigure}{S\arabic{figure}}
\renewcommand{\thetable}{S\arabic{table}}

\makeatletter
\renewcommand\@biblabel[1]{#1.}
\makeatother

\newcommand{\be}[1]{\begin{eqnarray}  {\label{#1}}}
\newcommand{\ee}{\end{eqnarray}}

\begin{center}
	\textbf{{\LARGE Supplementary Information}}
	
	for
	
	\textbf{{\Large Evolution of atomic optical selection rules upon gradual symmetry lowering}}
	
	by
	
	G.~J.~J.~Lof, C.~H.~van~der~Wal, and R.~W.~A.~Havenith
	
\end{center}

\vspace{4cm}

\noindent \textbf{TABLE OF CONTENTS}

\begin{figure}[h!]
	\includegraphics[width=\columnwidth]{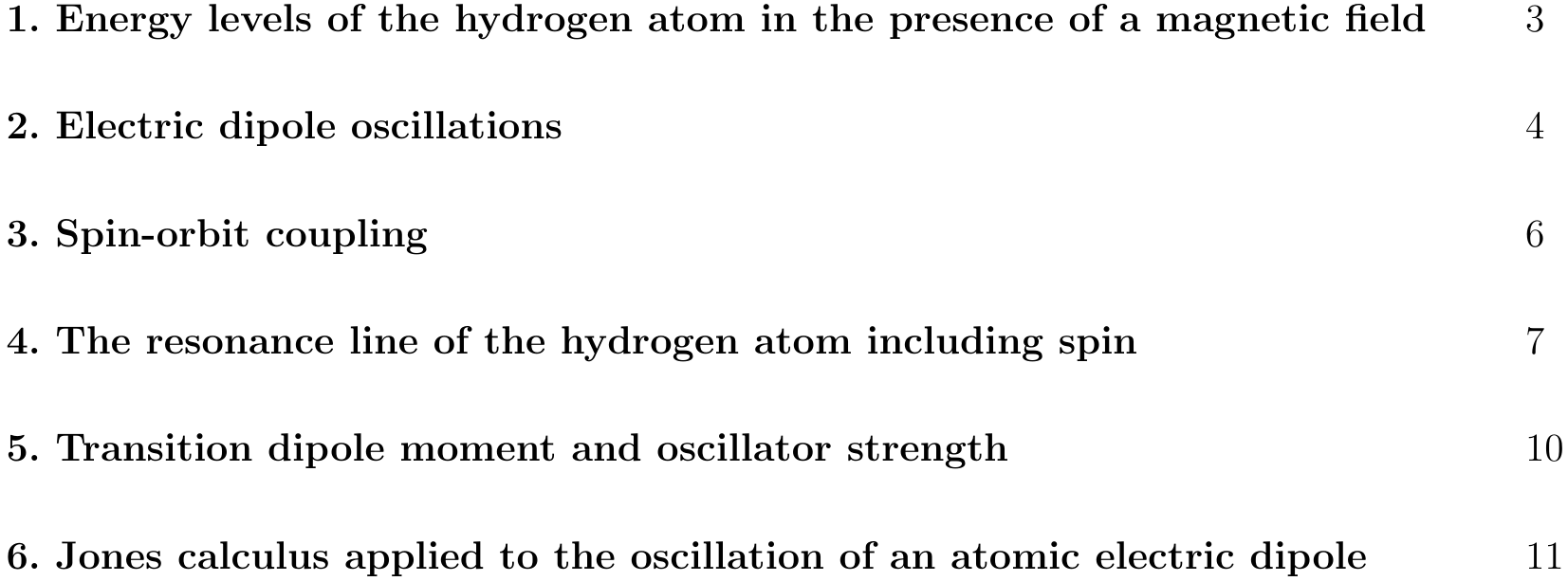}
\end{figure}

\newpage

\noindent In the main text we have considered the resonance lines of the hydrogen atom around 120 nm,
corresponding to an atomic transition between the ground state $|1s\rangle$ ($n = 1$; $l = m_l = 0$) and the excited state $|2p\rangle$ ($n = 2$; $l = 1$; $m_l = -1,0,1$). We have investigated the modification of the optical selection rules in the presence of surrounding negative point charges. Relevant fundamentals are given for reference in the Supplementary Information below. Analogous to the book of Cohen-Tannoudji, Diu and Lalo\"{e}\cite{cohen1991quantum} (which gives a derivation for the spinless case), the optical selection rules are derived for electronic transitions between $1s$ and $2p$ levels in the presence of a magnetic field while considering spin-orbit coupling, with the result tabulated in Table~\ref{table:288}. With these optical selection rules as a starting point (Fig.~\ref{fig:Transitions}b), we introduce point charges in the main text and study how the optical selection rules are modified.

\section{Energy levels of the hydrogen atom in the presence of a magnetic field}\label{Sec:SOM_Bfield}

For the hydrogen atom at zero magnetic field, the Hamiltonian $H_0$ (containing the kinetic and electrostatic interaction energy) has energy eigenvalues $E_n = -E_I/n^2$, with $E_I$ the ionization energy. In the presence of a static magnetic field $\textbf{B}$ along $z$, the resonance line is modified. This field does not only change the frequency, but also the polarization of the atomic lines, which is called the Zeeman effect. In addition, due to the electron and proton spins, the resonance line is affected by the fine- and hyperfine structure. However, let us neglect spin for the moment (in Section~\ref{Sec:SOM_Bfield} and~\ref{Sec:SOM_Dip}), following \cite{cohen1991quantum}.
As such, the Hamiltonian is given by $H = H_0 + H_1$, with $H_1$ the paramagnetic coupling term. The corresponding eigenvalue equation becomes
\begin{equation} 
(H_0 + H_1) |\phi_{n,l,m_l} \rangle = (E_n - m_l \mu_B B)) |\phi_{n,l,m_l} \rangle
\end{equation}
with $\mu_B = \frac{e \hbar}{2 m_e}$ the Bohr magneton, $e$ the elementary charge, $\hbar$ the reduced Planck constant, and $m_e$ the electron mass. For the states involved in the resonance line, we obtain
\begin{equation} 
(H_0 + H_1) |\phi_{1,0,0} \rangle = -E_I |\phi_{1,0,0} \rangle
\end{equation}
\begin{equation} 
(H_0 + H_1) |\phi_{2,1,m_l} \rangle = (-E_I  + \hbar(\Omega + m_l \omega_L)) |\phi_{2,1,m_l} \rangle
\end{equation}
with the Larmor angular velocity given by $\omega_L = \frac{e B}{2 m_e}$. At zero field this gives the angular frequency of the resonance line
\begin{equation} 
\Omega = \frac{E_2-E_1}{\hbar} = \frac{3 E_I}{4\hbar}
\end{equation}

\section{Electric dipole oscillations}\label{Sec:SOM_Dip}

The electric dipole operator is given by
\begin{equation} 
\textbf{D} = e \textbf{R}
\end{equation}
with $\textbf{R}$ the position operator. Hence, $\textbf{D}$ is a three-dimensional vector, with components $D_x$, $D_y$, $D_z$. Considering the $|1s\rangle$ and $|2p\rangle$ states of the hydrogen atom, the only non-zero matrix components of $\textbf{D}$ are\cite{cohen1991quantum}
\begin{equation} 
\begin{aligned}
\langle \phi_{2,1,1} | D_x |\phi_{1,0,0} \rangle &= -\langle \phi_{2,1,-1} | D_x |\phi_{1,0,0} \rangle = -\frac{e I_R}{\sqrt{6}} \\
\langle \phi_{2,1,1} | D_y |\phi_{1,0,0} \rangle &= \langle \phi_{2,1,-1} | D_y |\phi_{1,0,0} \rangle = i \frac{e I_R}{\sqrt{6}} \\
\langle \phi_{2,1,0} | D_z |\phi_{1,0,0} \rangle &= \frac{e I_R}{\sqrt{3}}
\end{aligned}
\end{equation}
where the constant value $I_R$ is a radial integral. If a system is in a stationary state, the mean value of the operator $\textbf{D}$ is zero, i.e.~the system cannot emit any light. Let us therefore assume that the system is in a linear superposition of the ground state $|1s\rangle$ and one of the $|2p\rangle$ excited state sublevels
\begin{equation} 
|\psi_{m_l} (t=0) \rangle = \cos(\alpha) |\phi_{1,0,0} \rangle + \sin(\alpha) |\phi_{2,1,m_l} \rangle
\end{equation}
with $\alpha$ real. As a function of time, this state evolves as
\begin{equation} 
|\psi_{m_l} (t) \rangle = \cos(\alpha) |\phi_{1,0,0} \rangle + \sin(\alpha) \mathrm{e}^{-i (\Omega+m \omega_L)t} |\phi_{2,1,m_l} \rangle
\end{equation}
where the global phase factor $\mathrm{e}^{-i E_i t /\hbar}$ has been omitted. The mean value of the electric dipole is given by
\begin{equation} 
\expval{\textbf{D}}_{m_l}(t) = \langle \psi_{m_l} (t) | \textbf{D} |\psi_{m_l} (t) \rangle
\end{equation}

For $m_l=1$, we obtain
\begin{equation} 
\begin{aligned}
\expval{D_x}_1 &= -\frac{e I_R}{\sqrt{6}} \sin(2\alpha) \cos((\Omega + \omega_L)t) \\
\expval{D_y}_1 &= -\frac{e I_R}{\sqrt{6}} \sin(2\alpha) \sin((\Omega + \omega_L)t) \\
\expval{D_z}_1 &= 0
\end{aligned}
\end{equation}
which implies that $\expval{\textbf{D}}_{1}(t)$ rotates in the $xy$-plane in the counter-clockwise direction, with angular velocity $\Omega + \omega_L$.

For $m_l=0$, we obtain
\begin{equation} 
\begin{aligned}
\expval{D_x}_0 &= \expval{D_y}_0 = 0 \\
\expval{D_z}_0 &=  \frac{e I_R}{\sqrt{3}} \sin(2\alpha) \cos(\Omega t)
\end{aligned}
\end{equation}
which implies that $\expval{\textbf{D}}_{0}(t)$ oscillates linearly along $z$, with angular frequency $\Omega$.

For $m_l=-1$, we obtain
\begin{equation} 
\begin{aligned}
\expval{D_x}_{-1} &= \frac{e I_R}{\sqrt{6}} \sin(2\alpha) \cos((\Omega - \omega_L)t) \\
\expval{D_y}_{-1} &= -\frac{e I_R}{\sqrt{6}} \sin(2\alpha) \sin((\Omega - \omega_L)t) \\
\expval{D_z}_{-1} &= 0
\end{aligned}
\end{equation}
which implies that $\expval{\textbf{D}}_{-1}(t)$ rotates in the $xy$-plane in the clockwise direction, with angular velocity $\Omega - \omega_L$.

For all three cases ($m_l = -1,0,1$) the mean value of the electric dipole oscillates as a function of time, corresponding to the emission of electromagnetic energy. The type of electric dipole oscillation determines the type of polarization of the emitted radiation. Still, the polarization of light that an observer sees depends on its orientation with respect to the source. For the $m_l = 1$ case, the electric dipole oscillates in the counter-clockwise direction with respect to the $z$-axis. An observer will at the positive (negative) side of the $z$-axis therefore detect $\sigma^+$ ($\sigma^-$) radiation, where $\sigma^{\pm} = \frac{x\pm iy}{\sqrt{2}}$.
However, if the observer detects in the $xy$-plane, the radiation will be linearly polarized, perpendicular to $\textbf{B}$. In any other direction, the radiation is elliptically polarized. For the $m_l = -1$ case, an observer will detect the opposite direction for circular and elliptical polarization. For the $m_l = 0$ case, an observer in the $z$-direction will not observe any radiation, since an oscillating linear dipole does not radiate along its axis. In any other direction the detected radiation will be linearly polarized, parallel to $\textbf{B}$.

If one is interested in excitation by means of polarized light, the process just takes place in the
reverse direction. Upon excitation, the electric dipole oscillation will become resonant to the oscillation of the electromagnetic field of a photon, where the polarization of the electric dipole oscillation is determined by the polarization of the photon. The type of dipole oscillation that is induced will be exactly the same as the type of oscillation that would be responsible for emission of this light. For example, if at $+z$ you observe $\sigma^-$ light induced by a clockwise rotation of the dipole (as seen from $+z$), you should excite the system with $\sigma^-$ with the source of light being positioned at $-z$, in order to induce the same oscillation (now counter-clockwise as seen from the origin).

\section{Spin-orbit coupling}\label{Sec:SOM_SOC}

According to special relativity, an electron moving in the electrostatic field of a proton experiences this field in its reference frame as a magnetic field\cite{atkins2011molecular}. The intrinsic magnetic moment due to the electron spin can interact with this magnetic field. The corresponding interaction energy is found to be proportional to the inner product of L and S, i.e.
\begin{equation} 
H_{SO} \propto \textbf{L} \cdot \textbf{S}
\end{equation}
Including this spin-orbit coupling (SOC), one obtains the total Hamiltonian
\begin{equation} 
H=H_0+H_{SO}
\end{equation}
with $H_0$ the original Hamiltonian without the spin-orbit interaction.

It is useful to define the total angular momentum operator
\begin{equation} 
J^2 = L^2 + S^2 + 2\textbf{L} \cdot \textbf{S}
\end{equation}
which allows to write
\begin{equation} 
\textbf{L} \cdot \textbf{S} = \frac{1}{2} (J^2 - L^2 - S^2)
\end{equation}
where the corresponding energies are determined from
\begin{equation} 
\expval{\textbf{L} \cdot \textbf{S}} = \frac{1}{2} (\expval{J^2} - \expval{L^2} - \expval{S^2})=\frac{\hbar^2}{2} (j(j+1) - l(l+1) -s(s+1))
\end{equation}
which implies that the spin-orbit interaction induces an energy splitting
\begin{equation} 
\Delta E \propto j(j+1) - l(l+1) -s(s+1)
\end{equation}
For atoms, the proportionality constant is
proportional to $Z^4$, with $Z$ the atomic number\cite{atkins2011molecular}. Hence, spin-orbit interaction is strong for (systems consisting of) heavy atoms.

\section{The resonance line of the hydrogen atom including spin}\label{Sec:SOM_Spin}

The electron spin can be either up ($\expval{S_z}=\hbar/2$) or down ($\expval{S_z}=-\hbar/2$), to which we refer as $\alpha$ and $\beta$, respectively. The orbitals
$1s$, $2p_{-1}$, $2p_{0}$ and $2p_{1}$ allow for eight possible spinorbitals. These spinorbitals span a basis to which we refer as the uncoupled representation\cite{cohen1991quantum}. It is convenient to label the basis states with the quantum numbers $l$, $s$, $m_l$ and $m_s$, as tabulated in Table~\ref{table:266}.

SOC lifts the 6-fold degeneracy of the $2p$ levels into sublevels with quantum number $j$, i.e.~$2p_{1/2}$ (2-fold degenerate) and $2p_{3/2}$ (4-fold degenerate).
The degeneracy can be further lifted by e.g.~a magnetic field (which introduces an additional term to $H$, which we neglect for the moment though), which induces a quantization axis. When the field is applied in the $z$-direction, it is convenient
to define a basis spanned by the eigenstates of the total angular momentum $J_z$. Constructing the matrix $J_z = L_z + S_z$ in the basis in which $H$ is diagonal, one finds that $J_z$ is block-diagonal, i.e. the basis does not consist of eigenstates of $J_z$. Diagonalization of the 2- and 4-fold degenerate subspaces provides the basis to which we refer as the coupled representation
(where $H$ remains diagonal).

When the spin and orbital angular momentum are coupled, it is convenient to work in the coupled representation (Table~\ref{table:277}).
Good quantum numbers are now $j$, $m_j$, $l$ and $s$. The basis can still be expressed as a linear combination of the basis states of the uncoupled representation, where the prefactors are the so-called Clebsch-Gordan coefficients.

Excitation or emission between a $1s$ and $2p$ sublevel is possible if a nonzero value is obtained for the transition dipole moment $\langle \psi_{e} | \textbf{D} |\psi_{g} \rangle$, with $|\psi_{g} \rangle = |j=\frac{1}{2},m_j=\pm \frac{1}{2},l=0,s=\frac{1}{2} \rangle$ and $|\psi_{e} \rangle = |j,m_j,l=1,s=\frac{1}{2} \rangle$.
As tabulated in Table~\ref{table:288}, we see that now ten of the twelve possible transitions from the $1s$ to $2p$ sublevels
have nonzero oscillator strength. Instead, in the case without SOC only six
transitions are possible (three for either up or down spin).

\begin{table}[h!]
	\centering
	\caption{
		\textbf{Uncoupled representation for the spinorbitals orginating from the $1s$, $2p_{-1}$, $2p_{0}$ and $2p_{1}$ orbitals.}}
	\begin{tabular}{|c | c|}
		\hline
		Spinorbital &	$|l,s,m_l,m_s \rangle$ \\ [0.5ex]
		\hline
		$|s\beta\rangle$ & $|0,\frac{1}{2},0,-\frac{1}{2} \rangle$ \\
		$|s\alpha\rangle$ & $|0,\frac{1}{2},0,\frac{1}{2} \rangle$ \\
		$|p_{-1}\beta\rangle$ & $|1,\frac{1}{2},-1,-\frac{1}{2} \rangle$ \\
		$|p_{-1}\alpha\rangle$ & $|1,\frac{1}{2},-1,\frac{1}{2} \rangle$ \\
		$|p_{0}\beta\rangle$ & $|1,\frac{1}{2},0,-\frac{1}{2} \rangle$ \\
		$|p_{0}\alpha\rangle$ & $|1,\frac{1}{2},0,\frac{1}{2} \rangle$ \\
		$|p_{1}\beta\rangle$ & $|1,\frac{1}{2},1,-\frac{1}{2} \rangle$ \\
		$|p_{1}\alpha\rangle$ & $|1,\frac{1}{2},1,\frac{1}{2} \rangle$ \\ [1ex]	
		\hline
	\end{tabular}
	\label{table:266}
\end{table}

\begin{table}[h!]
	\centering
	\caption{
		\textbf{Coupled representation for the
			spinorbitals originating from the $1s$, $2p_{-1}$, $2p_{0}$ and $2p_{1}$ orbitals.} The left column contains the basis states of the coupled representation. The middle column gives the same states in the basis of the uncoupled representation, where the prefactors are the so-called Clebsch-Gordan coefficients. The right column gives the notation as used in the main text, (for the excited state) corresponding to the case where no surrounding charges are present.}
	\begin{tabular}{|c | c| c|}
		\hline
		$|j,m_j,l,s \rangle$ &	$\sum C_{m_l,m_s} |l,s,m_l,m_s \rangle$ & \\ [0.5ex]
		\hline
		$|\frac{1}{2},-\frac{1}{2},0,\frac{1}{2} \rangle$ & $|0,\frac{1}{2},0,-\frac{1}{2} \rangle$ & $|g_1 \rangle$ \\
		$|\frac{1}{2},\frac{1}{2},0,\frac{1}{2} \rangle$ & $|0,\frac{1}{2},0,\frac{1}{2} \rangle$ & $|g_2 \rangle$ \\
		$|\frac{1}{2},-\frac{1}{2},1,\frac{1}{2} \rangle$ & $\frac{1}{\sqrt{3}} |1,\frac{1}{2},0,-\frac{1}{2} \rangle - \sqrt{\frac{2}{3}} |1,\frac{1}{2},-1,\frac{1}{2} \rangle$ & $|e_1 (q_0) \rangle$ \\
		$|\frac{1}{2},\frac{1}{2},1,\frac{1}{2} \rangle$ & $\sqrt{\frac{2}{3}} |1,\frac{1}{2},1,-\frac{1}{2} \rangle - {\frac{1}{\sqrt{3}}} |1,\frac{1}{2},0,\frac{1}{2} \rangle$ & $|e_2 (q_0) \rangle$ \\
		$|\frac{3}{2},-\frac{3}{2},1,\frac{1}{2} \rangle$ & $|1,\frac{1}{2},-1,-\frac{1}{2} \rangle$ & $|e_3 (q_0) \rangle$ \\
		$|\frac{3}{2},-\frac{1}{2},1,\frac{1}{2} \rangle$ & ${\frac{1}{\sqrt{3}}} |1,\frac{1}{2},-1,\frac{1}{2} \rangle + \sqrt{\frac{2}{3}} |1,\frac{1}{2},0,-\frac{1}{2} \rangle$ & $|e_4 (q_0) \rangle$ \\
		$|\frac{3}{2},\frac{1}{2},1,\frac{1}{2} \rangle$ & $\sqrt{\frac{2}{3}} |1,\frac{1}{2},0,\frac{1}{2} \rangle + {\frac{1}{\sqrt{3}}} |1,\frac{1}{2},1,-\frac{1}{2} \rangle$ & $|e_5 (q_0) \rangle$ \\
		$|\frac{3}{2},\frac{3}{2},1,\frac{1}{2} \rangle$ & $|1,\frac{1}{2},1,\frac{1}{2} \rangle$ & $|e_6 (q_0) \rangle$ \\ [1ex]	
		\hline
	\end{tabular}
	\label{table:277}
\end{table}

\begin{table}[t]
	\centering
	\caption{\textbf{Transition dipole moments, polarizations (pol.)~and transition strengths for transitions between $1s$ and $2p$ sublevels, in the presence of a magnetic field.}
		See main text Fig.~\ref{fig:Transitions} for definitions of
		$|g_\mu\rangle$ and $|e_\nu\rangle$.
		The relative oscillator strengths $f_{rel}$ (Eq.~\ref{eq:fRel}) for all transitions from either $|s_{-\frac{1}{2}}\rangle$ or $|s_{\frac{1}{2}}\rangle$ add up to $f_{rel,tot} = 9$, equal to the case without SOC (main text Sec.~\ref{Sec:noSpin}).}
	\begin{tabular}{|c | c| c| c|}
		\hline
		\vtop{\hbox{\strut $|s_{m_j}\rangle \leftrightarrow |p_{j,m_j}\rangle$} \hbox{\strut = $|g_\mu\rangle \leftrightarrow |e_\nu\rangle$}} &	\vtop{\hbox{\strut $\langle e_\nu | D | g_\mu \rangle$ in the uncoupled representation.}\hbox{Vanishing terms are omitted.}} & Pol. &
		$f_{rel}$ \\ [0.5ex]
		\hline
		\vtop{\hbox{\strut $|s_{-\frac{1}{2}}\rangle \leftrightarrow |p_{\frac{1}{2},-\frac{1}{2}}\rangle$} \hbox{\strut $= |g_1\rangle \leftrightarrow |e_1\rangle$}} & \vtop{\hbox{\strut $\Big( \frac{1}{\sqrt{3}} \langle 1,\frac{1}{2},0,-\frac{1}{2} | - \sqrt{\frac{2}{3}} \langle 1,\frac{1}{2},-1,\frac{1}{2} | \Big) D |0,\frac{1}{2},0,-\frac{1}{2} \rangle $}\hbox{\strut	$=\frac{1}{\sqrt{3}} \langle \phi_{2,1,0} | D_z |\phi_{1,0,0} \rangle =  \frac{1}{3}e I_R$}} & $\pi$ & 1 \\
		
		\vtop{\hbox{\strut$|s_{-\frac{1}{2}}\rangle \leftrightarrow |p_{\frac{1}{2},\frac{1}{2}}\rangle$} \hbox{\strut $ = |g_1\rangle \leftrightarrow |e_2\rangle$}}  & \vtop{\hbox{\strut $\Big( \sqrt{\frac{2}{3}} \langle 1,\frac{1}{2},1,-\frac{1}{2} | - \frac{1}{\sqrt{3}} \langle 1,\frac{1}{2},0,\frac{1}{2} | \Big) D |0,\frac{1}{2},0,-\frac{1}{2} \rangle$}\hbox{\strut	$=\sqrt{\frac{2}{3}} \langle \phi_{2,1,1} |  \frac{D_x + iD_y}{\sqrt{2}}  |\phi_{1,0,0} \rangle =  -\frac{\sqrt{2}}{3}e I_R$}} & $\sigma^+$ & 2 \\
		
		\vtop{\hbox{\strut$|s_{-\frac{1}{2}}\rangle \leftrightarrow |p_{\frac{3}{2},-\frac{3}{2}}\rangle$} \hbox{\strut $ = |g_1\rangle \leftrightarrow |e_3\rangle$}}  & \vtop{\hbox{\strut $\langle 1,\frac{1}{2},-1,-\frac{1}{2} | D |0,\frac{1}{2},0,-\frac{1}{2} \rangle$}\hbox{\strut$= \langle \phi_{2,1,-1} |  \frac{D_x - iD_y}{\sqrt{2}}  |\phi_{1,0,0} \rangle =  \frac{1}{\sqrt{3}}e I_R$}} & $\sigma^-$ & 3  \\
		
		\vtop{\hbox{\strut$|s_{-\frac{1}{2}}\rangle \leftrightarrow |p_{\frac{3}{2},-\frac{1}{2}}\rangle$} \hbox{\strut $ = |g_1\rangle \leftrightarrow |e_4\rangle$}}  & \vtop{\hbox{\strut $\Big( \frac{1}{\sqrt{3}} \langle 1,\frac{1}{2},-1,\frac{1}{2} | + \sqrt{\frac{2}{3}} \langle 1,\frac{1}{2},0,-\frac{1}{2} | \Big) D |0,\frac{1}{2},0,-\frac{1}{2} \rangle$}\hbox{\strut$=\sqrt{\frac{2}{3}} \langle \phi_{2,1,0} | D_z |\phi_{1,0,0} \rangle =  \frac{\sqrt{2}}{3}e I_R$}} & $\pi$ & 2 \\
		
		\vtop{\hbox{\strut$|s_{-\frac{1}{2}}\rangle \leftrightarrow |p_{\frac{3}{2},\frac{1}{2}}\rangle$} \hbox{\strut $ = |g_1\rangle \leftrightarrow |e_5\rangle$}}  & \vtop{\hbox{\strut $\Big( \sqrt{\frac{2}{3}} \langle 1,\frac{1}{2},0,\frac{1}{2} | + \frac{1}{\sqrt{3}} \langle 1,\frac{1}{2},1,-\frac{1}{2} | \Big) D |0,\frac{1}{2},0,-\frac{1}{2} \rangle$}\hbox{\strut	$=\frac{1}{\sqrt{3}} \langle \phi_{2,1,1} |  \frac{D_x + iD_y}{\sqrt{2}}  |\phi_{1,0,0} \rangle =  -\frac{1}{3}e I_R$}} & $\sigma^+$ & 1 \\
		
		\vtop{\hbox{\strut$|s_{-\frac{1}{2}}\rangle \leftrightarrow |p_{\frac{3}{2},\frac{3}{2}}\rangle$} \hbox{\strut $ = |g_1\rangle \leftrightarrow |e_6\rangle$}}  & $\langle 1,\frac{1}{2},1,\frac{1}{2} | D |0,\frac{1}{2},0,-\frac{1}{2} \rangle=0$ &  & 0 \\
		
		\vtop{\hbox{\strut$|s_{\frac{1}{2}}\rangle \leftrightarrow |p_{\frac{1}{2},-\frac{1}{2}}\rangle$} \hbox{\strut $ = |g_2\rangle \leftrightarrow |e_1\rangle$}}  & \vtop{\hbox{\strut $\Big( \frac{1}{\sqrt{3}} \langle 1,\frac{1}{2},0,-\frac{1}{2} | - \sqrt{\frac{2}{3}} \langle 1,\frac{1}{2},-1,\frac{1}{2} | \Big) D |0,\frac{1}{2},0,\frac{1}{2} \rangle$}\hbox{\strut$= \langle \phi_{2,1,-1} |  \frac{D_x - iD_y}{\sqrt{2}}  |\phi_{1,0,0} \rangle =  -\frac{\sqrt{2}}{3}e I_R$}} & $\sigma^-$ & 2  \\
		
		\vtop{\hbox{\strut$|s_{\frac{1}{2}}\rangle \leftrightarrow |p_{\frac{1}{2},\frac{1}{2}}\rangle$} \hbox{\strut $= |g_2\rangle \leftrightarrow |e_2\rangle$}}  & \vtop{\hbox{\strut $\Big( \sqrt{\frac{2}{3}} \langle 1,\frac{1}{2},1,-\frac{1}{2} | - \frac{1}{\sqrt{3}} \langle 1,\frac{1}{2},0,\frac{1}{2} | \Big) D |0,\frac{1}{2},0,\frac{1}{2} \rangle$}\hbox{\strut$=-\frac{1}{\sqrt{3}} \langle \phi_{2,1,0} | D_z |\phi_{1,0,0} \rangle =  -\frac{1}{3}e I_R$}} & $\pi$ & 1 \\
		
		\vtop{\hbox{\strut$|s_{\frac{1}{2}}\rangle \leftrightarrow |p_{\frac{3}{2},-\frac{3}{2}}\rangle$} \hbox{\strut $= |g_2\rangle \leftrightarrow |e_3\rangle$}}  &  $\langle 1,\frac{1}{2},-1,-\frac{1}{2} | D |0,\frac{1}{2},0,\frac{1}{2} \rangle =0$ &  & 0  \\
		
		\vtop{\hbox{\strut$|s_{\frac{1}{2}}\rangle \leftrightarrow |p_{\frac{3}{2},-\frac{1}{2}}\rangle$} \hbox{\strut $= |g_2\rangle \leftrightarrow |e_4\rangle$}}  & \vtop{\hbox{\strut $\Big( \frac{1}{\sqrt{3}} \langle 1,\frac{1}{2},-1,\frac{1}{2} | + \sqrt{\frac{2}{3}} \langle 1,\frac{1}{2},0,-\frac{1}{2} | \Big) D |0,\frac{1}{2},0,\frac{1}{2} \rangle$}\hbox{\strut$= \frac{1}{\sqrt{3}} \langle \phi_{2,1,-1} |  \frac{D_x - iD_y}{\sqrt{2}}  |\phi_{1,0,0} \rangle =  \frac{1}{3}e I_R$}} & $\sigma^-$ & 1 \\
		
		\vtop{\hbox{\strut$|s_{\frac{1}{2}}\rangle \leftrightarrow |p_{\frac{3}{2},\frac{1}{2}}\rangle$} \hbox{\strut $ = |g_2\rangle \leftrightarrow |e_5\rangle$}}  & \vtop{\hbox{\strut $\Big( \sqrt{\frac{2}{3}} \langle 1,\frac{1}{2},0,\frac{1}{2} | + \frac{1}{\sqrt{3}} \langle 1,\frac{1}{2},1,-\frac{1}{2} | \Big) D |0,\frac{1}{2},0,\frac{1}{2} \rangle$}\hbox{\strut$=\sqrt{\frac{2}{3}} \langle \phi_{2,1,0} | D_z |\phi_{1,0,0} \rangle =  \frac{\sqrt{2}}{3}e I_R$}} & $\pi$ & 2 \\
		
		\vtop{\hbox{\strut$|s_{\frac{1}{2}}\rangle \leftrightarrow |p_{\frac{3}{2},\frac{3}{2}}\rangle$} \hbox{\strut $ = |g_2\rangle \leftrightarrow |e_6\rangle$}}  & \vtop{\hbox{\strut $\langle 1,\frac{1}{2},1,\frac{1}{2} | D |0,\frac{1}{2},0,\frac{1}{2} \rangle$}\hbox{\strut$= \langle \phi_{2,1,1} |  \frac{D_x + iD_y}{\sqrt{2}}  |\phi_{1,0,0} \rangle =  -\frac{1}{\sqrt{3}}e I_R$}} & $\sigma^+$ & 3 \\ [1ex]	
		\hline
	\end{tabular}
	\label{table:288}
\end{table}

\FloatBarrier

\section{Transition dipole moment and oscillator strength}\label{Sec:SOM_TDM}
For a system with $N$ states, the transition dipole moment related to a transition from the initial state $|\psi_I\rangle$ to the final state $|\psi_F\rangle$ is given by $\boldsymbol{\mu}_{FI}$, where $I\in \{1...N\}$ and $F \in \{1...N\}$.
Within a Cartesian coordinate system ($i = \{ x,y,z \}$), the corresponding components of this complex vector are given by
\begin{equation}\label{eq:TDM1}
\mu_{FI,i} = \langle \psi_F | D_i | \psi_I \rangle
\end{equation}
where $D_i$ is the $i$-component of the electric dipole operator $\textbf{D} = e \textbf{R}$, with $e$ the elementary charge and $\textbf{R}$ the position operator. The transition dipole moment is Hermitian, implying that $\mu_{IF,i} = \langle \psi_I | D_i | \psi_F \rangle =\mu_{FI,i}^*$, where the $*$ denotes the complex conjugate.

When we consider spinorbitals, the $1s$ and $2p$ contain 2 and 6 sublevels, respectively. The total basis set contains thus 8 spinorbitals, for which one can write down the matrix elements of the transition dipole moment. For each $i$-component, we obtain an $8 \times 8$ matrix.
Since we are only interested in $|1s\rangle \leftrightarrow |2p\rangle$ transitions, the lower left $6 \times 2$ submatrix contains all the information we need, i.e.~the ones with $I \in \{1,2\}$ (1s sublevels) and $F\in \{3,8\}$ (2p sublevels). This matrix contains the elements of the transition dipole moment related to transitions from a $1s$ to a $2p$ sublevel. The upper right $2 \times 6$ matrix is related to transitions from a $2p$ to a $1s$ sublevel and contains the complex conjugates.

Let us from now on merely focus on the $6 \times 2$ lower left submatrix of $\mu_{FI,i}$ and relabel the states (according to main text Fig.~\ref{fig:Transitions}b~and~c). We denote $\boldsymbol{\mu}_{\nu\mu}$ as the transition dipole moment related to a transition from a ground state sublevel $| g_{\mu} \rangle$ ($\mu \in \{1,2\}$) to an excited state sublevel $| e_{\nu} \rangle$ ($\nu \in \{1,6\}$). Within a Cartesian coordinate system ($i = \{ x,y,z \}$), the corresponding components of this complex vector are given by
\begin{equation}\label{eq:TDM2}
\mu_{\nu\mu,i} = \langle e_\nu | D_i | g_\mu \rangle
\end{equation}
where $D_i$ is the $i$-component of $\textbf{D} = e \textbf{R}$, with $e$ the elementary charge and $\textbf{R}$ the position operator.
For each $i$-component, the transition dipole moments are conveniently put into a $6 \times 2$ matrix with values given by Eq.~\ref{eq:TDM2}.

A convenient measure for the strength of a transition is the real-valued oscillator strength $f$, which is proportional to the absolute square of the transition dipole moment\cite{atkins2011molecular}.
In this work, we will consider only a small subset of all possible transitions in the hydrogen atom, i.e.~the $|1s\rangle \leftrightarrow |2p\rangle$ transitions, where we refer to the sublevel transitions as $| g_{\mu} \rangle \leftrightarrow |e_{\nu} \rangle$.
The oscillator strength related to such a transition is therefore proportional to the absolute square of the transition dipole moment $\boldsymbol{\mu}_{\nu\mu}$ (which is a vector, such that we have to take the sum of the absolute squares of the components), i.e.
\begin{equation}\label{eq:fGen}
f_{\mu\nu} \propto |\boldsymbol{\mu}_{\nu\mu}|^2 = \sum_{i=x,y,z}|\mu_{\nu\mu,i}|^2
\end{equation}
where $f_{\mu\nu} = f_{\nu\mu}$. For our work it is convenient to define the relative oscillator strength related to a transition between a ground state $| g_{\mu} \rangle$ and an excited state $| e_{\nu} \rangle$ as
\begin{equation}\label{eq:fRel}
f_{rel,\mu\nu} = \frac{9}{(eI_R)^2} |\boldsymbol{\mu}_{\nu\mu}|^2 = \frac{9}{(eI_R)^2} \sum_{i=x,y,z}|\mu_{\nu\mu,i}|^2 =\frac{9}{(eI_R)^2} \sum\limits_{i=x,y,z}| \langle e_{\nu} | D_i | g_{\mu} \rangle |^2
\end{equation}
For an electron occupying the ground state sublevel $|g_{\mu}\rangle$, we define the total relative oscillator strength as
\begin{equation}\label{eq:fTot}
f_{rel,g_{\mu},tot} = \sum\limits_{\nu} f_{rel,\mu\nu} = \frac{9}{(eI_R)^2} \sum\limits_{\nu,i}| \langle e_{\nu} | D_i | g_{\mu} \rangle |^2
\end{equation}

\section{Jones calculus applied to the oscillation of an atomic electric dipole}\label{Sec:SOM_Jones}

To describe how polarized light is affected by interaction with an optical element (or a sample), it is often convenient to use Jones calculus. Within this method, light is represented by a Jones vector and the optical element by a Jones matrix. Within the Jones-vector formulation, a Jones vector contains the amplitude and phase of the electric field components of a light beam (orthogonal to its propagation direction). Commonly, the amplitudes are normalized, such that their intensities add up to 1. Any elliptical polarization can be described, including the special cases of linear and circular polarization.

The polarization ellipse is described by the azimuth $\theta$ and the ellipticity angle $\epsilon$, as illustrated in main text Fig.~\ref{fig:PolarizationEllipse}. The azimuth $\theta$ is the angle between the semi-major axis $a$ and the horizontal axis, where $-\frac{1}{2} \pi \leqslant \theta < \frac{1}{2} \pi$. Note that $a$ and $\theta$ are ill-defined for circularly polarized light. The ellipticity angle $\epsilon$ is defined through the ellipticity $e = \frac{b}{a}$ (with $b$ the semi-minor axis) such that $e = \pm \tan \epsilon$, where $-\frac{1}{4} \pi \leqslant \epsilon \leqslant \frac{1}{4} \pi$ (where the $+$ and $-$ signs correspond to right- and left-handed polarization respectively).

The general definition of the Jones vector representing an electric vector oscillating in the $xy$-plane is given by\cite{r1987ellipsometry}
\begin{equation}\label{eq:Jones00}
\textbf{E}\{\hat{\textbf{x}},\hat{\textbf{y}}\}=
A \mathrm{e}^{i \delta} \textbf{R}(-\theta)
\left[
\begin{matrix}
\cos(\epsilon) \\ i \sin(\epsilon)
\end{matrix}
\right]  = A \mathrm{e}^{i \delta}
\left[
\begin{matrix}
\cos(\theta) \cos(\epsilon) - i \sin(\theta) \sin(\epsilon)\\ \sin(\theta) \cos(\epsilon) + i \cos(\theta) \sin(\epsilon)
\end{matrix}
\right]
\end{equation}
where we will take for convenience the amplitude $A=1$ and the global phase $\delta=0$. The transformation matrix $\textbf{R}(-\theta)$ rotates the primed basis with an angle $-\theta$ (to the unprimed basis) in main text Fig.~\ref{fig:PolarizationEllipse}. It turns out that we have to distinguish only two cases in this work, i.e.~$\theta = 0$ (the $x$-axis is the major axis) and $\theta = \pi/2$ (the $y$-axis is the major axis). Within the $\{\hat{\textbf{x}},\hat{\textbf{y}}\}$-basis, the corresponding Cartesian Jones vectors follow from Eq. (\ref{eq:Jones00}) by substituting for $\theta$, from which we can easily find $\epsilon$, i.e.
\begin{equation}\label{eq:Jones11}
\theta=0: \quad  \textbf{E}\{\hat{\textbf{x}},\hat{\textbf{y}}\}=
\left[
\begin{matrix}
\cos(\epsilon) \\ i \sin(\epsilon)
\end{matrix}
\right], \quad
\epsilon = \sin^{-1}(-iE_y)
\end{equation}
\begin{equation}\label{eq:Jones12}
\theta=\pi/2: \quad  \textbf{E}\{\hat{\textbf{x}},\hat{\textbf{y}}\}=
\left[
\begin{matrix}
-i\sin(\epsilon) \\ \cos(\epsilon)
\end{matrix}
\right], \quad
\epsilon = \sin^{-1}(iE_x)
\end{equation}
which for both cases corresponds to a clockwise rotation (right-handed polarization) for $\epsilon>0$. Note that $|E_x|>|E_y|$ when $\theta = 0$, whereas $|E_y|>|E_x|$ when $\theta = \pi/2$.

We find it convenient to assign a Jones vector to
the oscillation of an atomic electric dipole related to an electronic transition, with the components of the electric vector $\textbf{E}$ given by the (normalized) components of the corresponding transition dipole moment. As such, an electric dipole oscillating in the $xy$-plane is represented by the Jones vector
\begin{equation} \label{eq:Evector}
\textbf{E}\{\hat{\textbf{x}},\hat{\textbf{y}}\}= N
\left[
\begin{matrix}
\langle \phi_f | D_x | \phi_i \rangle \\ \langle \phi_f | D_y | \phi_i \rangle
\end{matrix}
\right]
\end{equation}
with $N$ a normalization constant and $| \phi_{i(f)} \rangle$ the initial (final) state. We will always take the ground state $| \phi_{f} \rangle = | g_{1,2} \rangle$ for the initial state.

For the $1s \leftrightarrow 2s$ transitions of the hydrogen atom, it turns out that for six of the twelve possible transitions the transition dipole moment has only nonzero $z$-components for increasing charge $q$, i.e.~the electric dipole oscillates only in the $z$-direction (these six have been named Group Z (blue) in main text Table~\ref{table:pola}). For this Group Z, only the oscillator strength varies as a function of $q$ (i.e.~the polarization selection rules vary only in the sense of varying between forbidden and allowed). For the other six transitions (Group XY (red) in main text Table~\ref{table:pola}), the transition dipole moment has only nonzero $x$- and $y$-components, i.e.~the electric dipole oscillates in the $xy$-plane. For these transitions the polarization seletion rules are also affected in the sense of a change of the ellipticity (actually, only the ellipticity angle $\epsilon$ turns out to be affected).
We find for this Group XY for increasing charge $q$ that of the components $E_x$ and $E_y$ always one is purely real and the other imaginary (or zero). This implies that we have either $\theta=0$ or $\theta=\pi/2$. More specific, for increasing $q$ the polarization selection rules for each of the Group XY series change from circular to linear, without affecting $\theta$.
Therefore, we can for each series always write the Jones vector in the form of either Eq.~(\ref{eq:Jones11}) or (\ref{eq:Jones12}), where we multiply with a global phase factor $\mathrm{e}^{i \delta'}$ with the phase $\delta'$ taken such that $E_x$ becomes real when $|E_x| > |E_y|$ and $E_y$ becomes real when $|E_y| > |E_x|$. Subsequently, we can easily determine the ellipticity angle $\epsilon$.

\end{document}